\documentclass[sigconf]{acmart}

\usepackage{enumitem}
\usepackage{natbib}  
\usepackage{hyperref}
\usepackage{graphicx}
\usepackage{amsmath}
\usepackage{multirow}
\usepackage{makecell}
\usepackage{subcaption}
\usepackage{float}
\usepackage{xcolor}

\AtBeginDocument{%
  }

\setcopyright{acmcopyright}
\copyrightyear{2023}

\begin{document}

\newcommand{\papertitle}{NIRVANA: A Comprehensive Dataset for Reproducing How Students Use Generative AI for Essay Writing}
\newcommand{\out}[1]{}
\newcommand{\sang}[1]{\out{{\small\textcolor{blue}{\bf [*** sang: #1]}}}}
\newcommand{\jelson}[1]{\out{{\small\textcolor{red}{\bf[*** Jelson: #1]}}}}

\newcommand{\rqone}{Research Question 1}

\newcommand{\datalinkanon}{\url{https://osf.io/3a8uh/overview?view_only=1f1af82b22384961b682d9cba5d1661f
}}

\title{\papertitle}

\author{Andrew Jelson}
\affiliation{%
  \institution{Virginia Tech}
  \city{Blacksburg}
  \state{Virginia}
  \country{USA}
  \postcode{24060}
}
\email{jelson9854@vt.edu}

\author{Daniel Manesh}
\affiliation{%
  \institution{Virginia Tech}
  \city{Blacksburg}
  \state{Virginia}
  \country{USA}
  \postcode{24060}
}
\email{danielmanesh@vt.edu}

\author{Sangwook Lee}
\affiliation{%
  \institution{Virginia Tech}
  \city{Blacksburg}
  \state{Virginia}
  \country{USA}
  \postcode{24060}
}
\email{sangwooklee@vt.edu}

\author{Alice Jang}
\affiliation{%
  \institution{Virginia Tech}
  \city{Blacksburg}
  \state{Virginia}
  \country{USA}
  \postcode{24060}
}
\email{ajjang@vt.edu}

\author{Daniel Dunlap}
\affiliation{%
  \institution{Virginia Tech}
  \city{Blacksburg}
  \state{Virginia}
  \country{USA}
  \postcode{24060}
}
\email{dunlapd@vt.edu}

\author{Tamara Maddox}
\affiliation{%
  \institution{George Mason University}
  \city{Fairfax}
  \state{Virginia}
  \country{USA}
  \postcode{22030}
}
\email{tmaddox@gmu.edu}

\author{Young-Ho Kim}
\email{yghokim@younghokim.net}
\affiliation{%
  \institution{NAVER AI Lab}
  \city{Seongnam}
  \country{South Korea}
}

\author{Sang Won Lee}
\authornote{Sang Won Lee conducted this work while at NAVER AI Lab as a visiting scholar.}
\email{sangwonlee@vt.edu}
\affiliation{%
  \institution{Virginia Tech}
  \city{Blacksburg}
  \state{Virginia}
  \country{USA}
  \postcode{24061}
}

\renewcommand{\shortauthors}{Jelson, Manesh, Sangwook Lee, Sang won Lee}

\begin{abstract}
With the rapid adoption of AI writing assistants in education, educators and researchers need empirical evidence to understand the impact on student writing and inform effective pedagogical design. Despite widespread use, we lack systematic understanding of how students engage with these tools during authentic writing tasks: when they seek assistance, what they ask, and how they incorporate AI-generated content into their essays. This gap limits evidence-based policy development and rigorous evaluation of generative AI's learning effects. To address this gap, we introduce NIRVANA, a dataset capturing how university students use generative AI while writing an analytical essay. The dataset includes 77 students who completed an essay task with access to ChatGPT, recording keystroke-level writing behavior, full ChatGPT conversation histories, and all text copied from ChatGPT, enabling a complete reconstruction of the writing process and revealing how AI assistance shapes student work. Our analysis identifies key behavioral patterns, including variation in ChatGPT query frequency and its relationship to essay characteristics such as length and readability. We identify four writing profiles based on students' contribution and revision patterns: \textit{Lead Authors, Collaborators, Drafters, and Vibe Writers}. To support deeper investigation, we developed a replay interface that reconstructs the writing process; qualitative analysis of sampled replays demonstrates how this tool enables systematic examination of student–AI interactions. This work contributes four resources: (1) a publicly available dataset of naturalistic student–AI writing interactions, (2) empirical insights into student engagment with AI writing assistants, (3) identification of four distinct writer profiles, and (4) a replay tool for analyzing temporal writing data. Together, these contributions enable future research on academic integrity, learning outcomes, and pedagogical design as generative AI reshapes educational practice.
\end{abstract}
% We anticipate that this dataset, with the replay interface, can provide a scalable approach for instructors and researchers examining AI-assisted writing as well as an in-depth trace that one can account for a writer's usage, if not intention. 



%%\received{20 February 2007}
%%\received[revised]{12 March 2009}
%%\received[accepted]{5 June 2009}

\maketitle

\section{Introduction}

AI writing assistants have been used for more than a decade to support various writing tasks, including revising sentences, suggesting word choices, translating text, and generating content~\cite{weber_legalwriter_2024, knight_acawriter_2020}. As these tools become increasingly advanced and widely accessible, they are reshaping the way students engage with academic writing. While some educators see AI as a powerful aid for improving writing quality and accessibility~\cite{bower_how_2024, barrett_not_2023}, others raise concerns about academic integrity, over-reliance, and the potential decline of students' critical thinking skills~\cite{kosmyna_your_nodate, adams_artificial_2022, cotton_chatting_2023}. The rise of ChatGPT and other large language models (LLMs) has intensified these debates, leading to calls for deeper investigation into student interactions with these tools~\cite{jelson_empirical_2025, liu_teaching_2024, bower_how_2024}.

Existing research has explored AI writing tools from various angles -- analyzing the quality of AI-generated text, investigating detection methods, and debating ethical implications~\cite{goldi_intelligent_2024, knight_acawriter_2020, weber_legalwriter_2024}. 
However, a critical gap remains: existing studies rely primarily on surveys and self-reports~\cite{wang_effect_2025, cotton_chatting_2023}, which cannot reveal the temporal dynamics and quantitative analysis at scale of AI-assisted writing: when students seek help, how they integrate AI-generated content, and how these interactions unfold throughout the writing process. Moreover, researchers lack validated methodological tools for analyzing such complex temporal data at scale.
To address these gaps, we present NIRVANA (Naturalistic Interactions and Replay of Voluntary AI-Assisted Nonfiction Academic Writing), a dataset and replay platform designed to investigate how students use generative AI when writing argumentative essays. We collected data from 77 students completing essay assignments with access to ChatGPT. Students wrote in a custom-built environment integrated with an in-house version of ChatGPT, which recorded the entire writing process at the keystroke level. The system captured user inputs, copy-and-paste events, and complete ChatGPT conversation histories with timestamps.

To illustrate the dataset's analytical potential, we conducted exploratory analyses to characterize patterns of AI usage and examine their relationships with writing outcomes. Finally, we developed a replay interface that allows researchers and educators to reconstruct and examine the writing process chronologically. A subset of essays was closely inspected to enable deeper observation of how AI use may have influenced students’ learning.

In conclusion, NIRVANA enables a scalable and in-depth approach to studying AI-assisted writing.
This paper makes four primary contributions:
\begin{enumerate}
\item NIRVANA dataset --- a publicly available dataset documenting authentic student-AI writing interactions with temporal granularity.
\item Quantitative analysis revealing patterns in AI usage and their correlations with essay characteristics, such as length, time taken, and readability.
\item Identification of four distinct writer profiles, characterized by different patterns in how they contribute to the essay.
\item NIRVANA application --- A replay interface that enables in-depth analysis of essay writing and generative AI use.
\end{enumerate}

Together, these resources provide a number of research communities --- including learning science, human-computer interaction, and natural language processing --- with empirical data and tools needed to understand and shape generative AI's role in educational and writing contexts.
% % Our dataset and analytical tools enable investigation of pedagogically important questions, including:
% % \begin{itemize}
% % \item What are the RQs?

% % \item How does the timing and frequency of AI assistance relate to writing qdata-collection methodology, describes the dataset's structure strategies indicate meaningful engagement with AI-generated content versus surface-level copying?
% % \item How do AI usage patterns differ across students and writing stages, and what implications does this have for scaffolding and instruction?
% \end{itemize}

This paper details our data-collection methodology, describes the dataset structure, and presents our exemplary quantitative and qualitative analysis of student-AI interactions that the dataset and the replay interface enable. We conclude with implications for educators and researchers and outline directions for future work.

\section{Related Work} 
%This part is under reconstruction

\subsection{AI Writing Assistance in Educational Contexts}

For years, AI has been incorporated into writing education through automated feedback systems and tutoring platforms~\cite{knight_acawriter_2020, afrin_effective_2021, crompton_artificial_2023} to support the learning process by helping students communicate ideas, fostering critical thinking, analysis, and synthesis~\cite{condon2004assessing, wade1995using, zinsser1989writing}. AI writing assistants can be used to improve writing process and support learning outcomes for students. Grammarly, a commercially available tool, helps students avoid plagiarism and improve their writing quality, especially for EFL learners~\cite{grammarly, koltovskaia_student_2020, xiong_automated_2019}, though these students often underestimate its capabilities~\cite{huang_effectiveness_2020}. Educational systems, like AcaWriter, demonstrate that essay writing can be supported by automated rhetorical feedback~\cite{knight_acawriter_2020}. RECIPE extends this by integrating ChatGPT into EFL classrooms, showing improved performance and satisfaction when given AI-mediated revision support~\cite{han_recipe_2023}. Similarly, CoachGPT provides scaffolding-based support for essay planning and reviewing, demonstrating that structured AI guidance can improve student writing outcomes and engagement~\cite{chen_coachgpt_2025}. LegalWriter shows LLM-based feedback improves writing quality and student learning outcomes, in interdisciplinary legal writing contexts~\cite{weber_legalwriter_2024}.

Observational studies of AI-assisted writing reveal consistent patterns in how both students and professional writers engage with these tools in practice. Research on students shows that they engage more when using ChatGPT for reviewing and revision processes~\cite{goldi_intelligent_2024}. LLM usage does not necessarily speed up the writing process; students report increased time spent on their work and perceive higher quality outcomes~\cite{goldi_intelligent_2024, wilbers_overall_2024}. HCI research has also examined how writers use AI more broadly, revealing that they typically seek assistance with translating ideas and reviewing drafts, with general appreciation for ChatGPT's ability to generate unexpected and sometimes inspiring output~\cite{chakrabarty_creativity_2024, gero_social_2023, lee_coauthor_2022}.

Recent research has examined how individual characteristics shape AI engagement in writing contexts, with implications into how educational systems might personalize or scaffold AI access. \citeauthor{joshi_writing_2025} finds that perceived ownership of AI-assisted essays is shaped by the length and detail of student prompts~\cite{joshi_writing_2025}, suggesting that prompting behavior is important for learner agency. Other work suggests that ownership perceptions are varied depending on the writing tasks~\cite{wasi_llms_2024}. Researchers have also identified how personal values, relationships with AI, and individual integration strategies shape student engagement with AI tools, highlighting the need to account for different student backgrounds~\cite{guo_pen_2025}.

\subsection{Writing Pedagogy and Educational Concerns}

Essay writing serves as a critical site for learning across disciplines, with feedback playing a critical role in developing both writing competencies and domain knowledge. Research in writing pedagogy has consistently demonstrated that timely, specific feedback improves not only the quality of student writing but also deeper conceptual understanding of subject matter~\cite{tan_reframing_2024, xiong_automated_2019}. Formative feedback --- delivered during the drafting process rather than solely on final products --- has been shown to support iterative revision practices that mirror expert writing behaviors~\cite{zhang_friction_2025}. The challenge in educational settings remains scalability: providing individualized feedback to large numbers of students requires significant time and effort, motivating interest in technological approaches that might augment or complement human feedback while preserving its pedagogical benefits.

Despite these promising findings, educators and practitioners have raised concerns about the negative impacts of generative AI that are difficult to regulate at scale, including weakened critical engagement, reduced cognitive effort, and threats to academic integrity~\cite{Livingstone2024, EKE2023100060, scarfe2024real}. Research also shows limits to AI writing quality, with AI-generated essays passing evaluation criteria 3–10 times less frequently than human-written essays~\cite{chakrabarty_art_2024, romoff_role_2025}. These concerns are particularly important in large-scale educational contexts, where instructors have limited visibility into how students are engaging with AI tools and whether that engagement supports or undermines learning. However, the majority of observational work focuses on professional or creative writers rather than students in authentic educational settings --- a critical gap, as students are learning writing skills rather than using AI in professional settings.

While this growing body of work provides valuable insights into AI adoption and writing processes, most observational studies focus on professional writers and do not make their data openly available for further analysis. There is a lack of publicly accessible datasets capturing how students engage with generative AI during authentic academic writing tasks, limiting the ability of researchers and educators to study these interactions. To address this gap, we contribute a dataset of student essays written with ChatGPT assistance, NIRVANA, alongside a replay system that allows researchers to reconstruct and analyze individual writing sessions.

\begin{figure*}[t]
    \centering
    \includegraphics[width=0.9\linewidth]{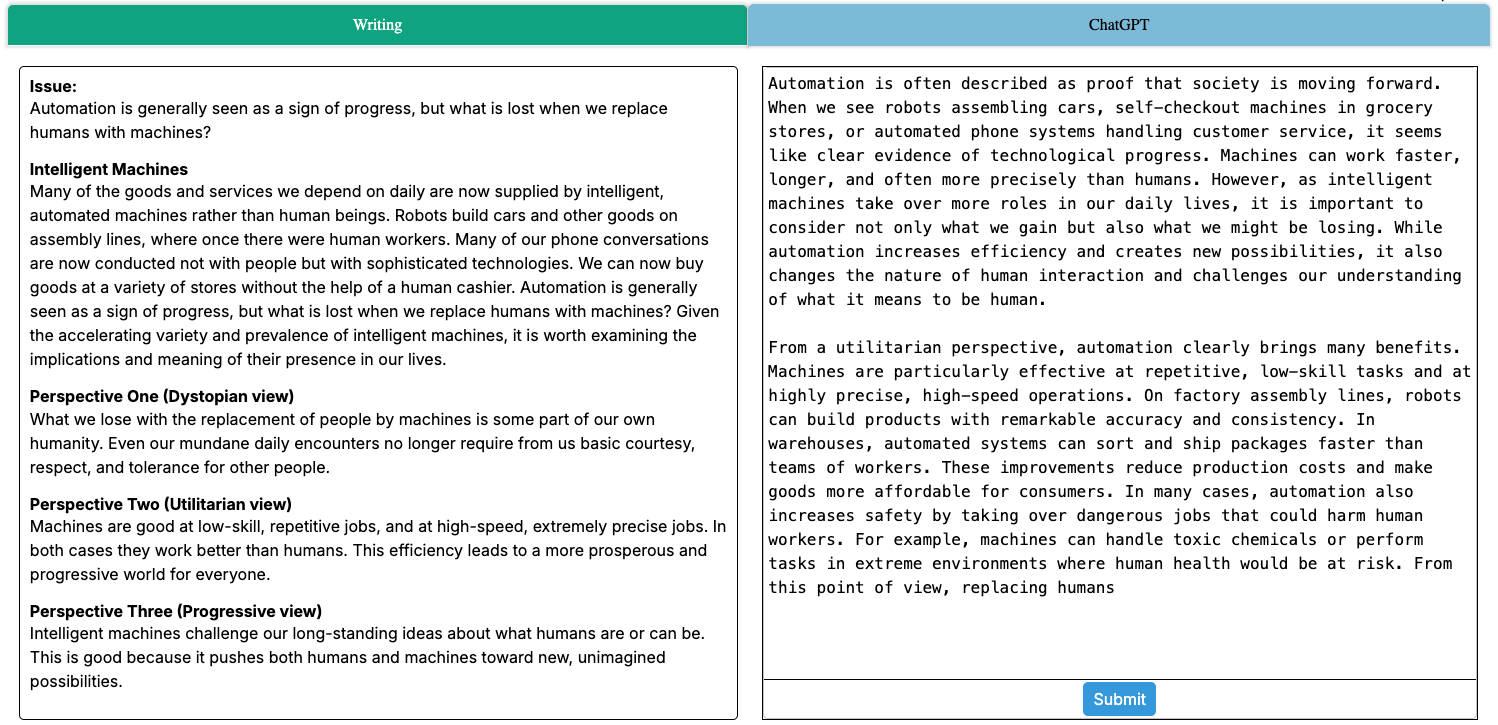}
    \caption{A screenshot of the Editor Interface}
    \label{fig:writ}
\end{figure*}

\section{Dataset Generation}

\subsection{Collection System Overview}
\label{sec:data-collection-system}
We developed a custom writing environment to generate a dataset that enables researchers to examine how students use generative AI tools incorporate AI-generated suggestions into their writing process. Because ChatGPT operates as an independent application that instructors typically cannot access or monitor, we integrated an in-house instance of ChatGPT directly into the writing environment. This integration allowed the system to systematically record user interactions, including students’ queries and the corresponding AI responses. 

The environment consists of two components: a text enditor and an in-house ChatGPT interface 
The text editor served as the primary writing interface, where participants composed essays in response to a given prompt (shown in \autoref{fig:writ}). To analyze the writing process, we logged all input events, including keystrokes, cursor movements, insertions, deletions, and copy–paste actions. 

The in-house ChatGPT interface was designed to mirror the functionality of the standard ChatGPT application, enabling users to engage in conversational exchanges with an AI assistant (see \autoref{fig:gpt} in \autoref{app:gpt}). To implement this functionality, we developed a custom ChatGPT system using the OpenAI API with the GPT-3.5-turbo model, which was the default version at the time of data collection. Participants were allowed to ask questions without restriction, and we did not provide a predefined system prompt (e.g., assigning the model a specific role such as a writing assistant), allowing its use to remain flexible and student-driven. 

Using this environment, we recorded three types of data: (1) students’ queries to ChatGPT, (2) ChatGPT’s responses, and (3) keystroke-level logs capturing the full writing process.Timestamps were recorded for each event, allowing us to identify when students sought assistance from ChatGPT and how they incorporated AI-generated content into the essay. The logging and replay functionalities were implemented using the CodeMirror 5 API and the CodeMirror-Record library~\cite{noauthor_jisuankecodemirror-record_2025}. All recorded data was stored locally on the user's machine and sent to our server upon submission of their essay.

% We recorded their queries and timestamps for analysis of when and how ChatGPT was prompted for assistance during the writing process. 

\subsection{Collection Process}
We administered pre- and post-study surveys to collect participant information and assess their writing experience. The pre-study survey, created in QuestionPro, gathered demographic data and included two validated instruments: an adapted Technology Acceptance Model (TAM) questionnaire for ChatGPT~\cite{davis_perceived_1989} and the Self-Efficacy for Writing Scale (SEWS)~\cite{bruning_examining_2013}. The measures allowed us to examine how their prior attitudes toward ChatGPT and their writing self-efficacy related to their usage patterns and writing.

\begin{figure*}[ht]
    \begin{subfigure}{0.49\textwidth}
        \centering
        \includegraphics[width=\textwidth]{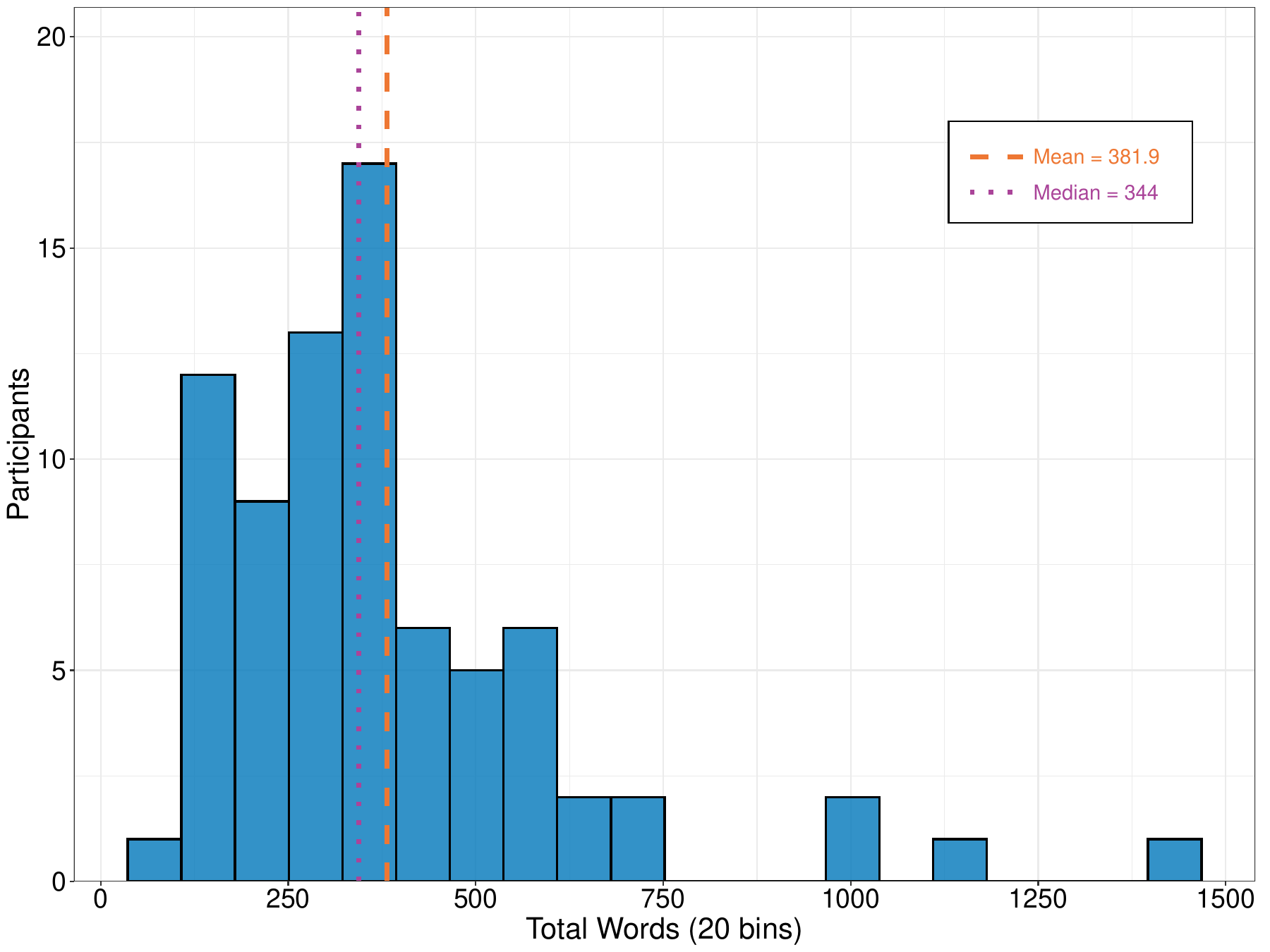}
        \caption{Word count distribution across participants}
        \label{fig:word}
    \end{subfigure}
    % \hfill
    \begin{subfigure}{0.49\textwidth}
        \centering
        \includegraphics[width=\textwidth]{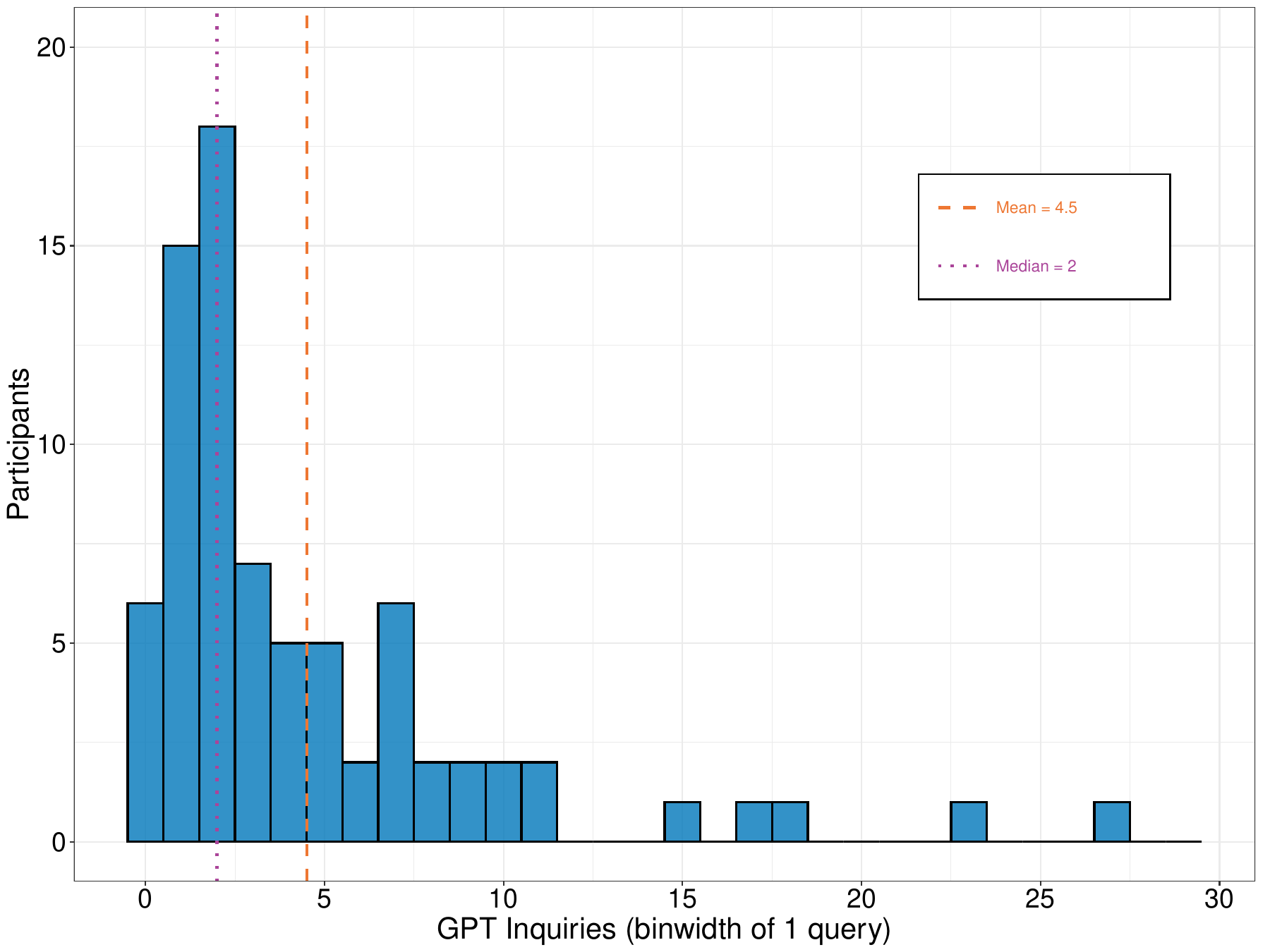}
        \caption{Inquiry distribution across participants}
        \label{fig:inq}
    \end{subfigure}
    \caption{Distribution of word count and inquiry counts across participants}
    \vspace{-10pt}
\end{figure*}

After submitting their essays, participants completed two additional questionnaires in the post-study survey to reflect on their writing experience. To measure whether students perceived the essay as truly "theirs" and how ChatGPT influenced that perception, we used a validated Perceived Ownership (PO) scale~\cite{avey_psychological_2009, chantal_psychological_2012, vandewalle_psychological_1995}.
Second, we administered the Creativity Support Index (CSI) to evaluate how effectively ChatGPT supported their creativity~\cite{cherry_quantifying_2014}. We excluded the Collaboration subscale, as the task did not involve interaction with other people. Together, these pre- and post-study measures enabled us to examine individual differences influencing ChatGPT use and how such use shaped students’ perceptions of their work.

\subsection{Participant Information}
Participants were recruited through university mailing lists and Prolific, an online crowdsourcing platform, with eligibility restricted to college students in the United States. All participants were entered into a raffle for a \$10 gift card (odds: 1 in 5). In total, 77 participants completed the study.
Participants’ ages were distributed as follows: 18–24 (n = 62), 25–34 (n = 10), 35–44 (n = 2), 45–54 (n = 1), and 55–64 (n = 2). Of the 77 participants, 34 identified as women, 1 as nonbinary, and 42 as men. The reported racial distribution was 37 White/Caucasian, 24 Asian/Pacific Islander, 7 Black or African American, 5 Hispanic, and 4 Other.

\subsection{Study Procedure and Task Information}
Before the writing task, participants completed a pre-study survey collecting demographic information and two validated measures: an adapted Technology Acceptance Model (TAM) scale for ChatGPT~\cite{davis_perceived_1989} and the Self-Efficacy for Writing Scale (SEWS)~\cite{bruning_examining_2013} (see \autoref{app:pre_survey}). TAM captures perceived usefulness and ease of use of the technology, whereas SEWS assesses confidence in writing ability. These measures, as part of the dataset, can be used to examine how beliefs about writing competence and perceptions of ChatGPT relate to their ChatGPT usage patterns.

Participants completed an essay task using our integrated writing platform with embedded ChatGPT. The prompt, adapted from an American College Testing (ACT) essay~\cite{act_prompt}, addressed the issue of automation replacing human labor and presented three perspectives. Participants were asked to develop their own position and relate it to at least one of the provided perspectives (\autoref{fig:writ}). They were advised to spend approximately 30 minutes on the task.
Although instructed to treat the essay as a graded class assignment and informed that they could use the in-house ChatGPT interface if they wished, participants retained full autonomy over how to use the tool. We imposed no minimum length requirement, and compensation was not contingent on essay quality or performance.

After submitting their essays, participants completed post-task measures assessing perceived ownership of the essay~\cite{avey_psychological_2009, chantal_psychological_2012, vandewalle_psychological_1995} and perceived creativity support using the Creativity Support Index (CSI)~\cite{cherry_quantifying_2014}, excluding the collaboration subscale. These measures capture students’ subjective experiences of authorship and creative support when writing with ChatGPT. 

\subsection{Dataset Structure}
Our dataset can be found at {\datalinkanon} as a spreadsheet file. We removed personally identifiable information and preprocessed the data to a more readable format. The dataset has two spreadsheets. The first contains participant information we collected from the pre- and post-study surveys. The second spreadsheet includes participants' essay data showing addition, deletion, cursor moves, copy-paste, and GPT query/response. A full list of data structures is available in \autoref{app:columns}.

\section{NIRVANA -- Dataset Overview}

We report descriptive statistics characterizing participants' writing processes and AI usage patterns. On average, participants produced $382$ words per essay ($\sigma=235$) and a median of $344$ words, indicating large variability in essay length (shown in \autoref{fig:word}). Participants typically spent $22.5$ minutes ($\sigma=15.10$) completing the task, with a median of $20.6$ minutes, indicating notable variability in task completion time. Additionally, participants made, on average, $4.51$ ChatGPT queries ($\sigma=5.11$) with a median of $2$ queries, indicating substantial variability in usage patterns. The query counts were highly right-skewed (skew = $2.24$) as shown in \autoref{fig:inq}, suggesting that while most participants consulted ChatGPT infrequently ($0-2$ times), a subset engaged extensively with the AI assistant.

We investigated the complexity of the participants' essays using the Dale-Chall readability score. The Dale-Chall score evaluates readability based on sentence length and word difficulty~\cite{dale_formula_1948}, and has been found to be more accurate than other commonly used readability metrics~\cite{klare_measurement_2000}, making it well-suited to assess the complexity of student writing. Participants averaged a Dale-Chall grade level of $9.96$ ($\sigma=1.34$) with a median of $10.07$, indicating that most essays were written at a college reading level (9.0+). The distribution was approximately symmetric (skew = $-0.08$, kurtosis = $-0.91$), suggesting that writing complexity was relatively consistent between participants, with scores evenly distributed around the mean.

% As part of the post-survey, we wanted to investigate ChatGPT's impact on the writing experience. The participants' perceived ownership scores averaged $5.3$ ($\sigma=1.33$) and had a median of $5.75$, indicating moderate variability. The distribution was moderate to highly left-skewed (skew = $-0.9$, kurtosis = $0.07$), indicating that most participants felt strong ownership of their essays with scores concentrated at the upper end, though a notable group reported lower ownership levels. On average, participants reported Creativity Support Index scores of $5.15$ ($\sigma=1.04$) with a median of $5.00$, indicating moderate variability. This data was approximately normally distributed (skew = $-0.26$, kurtosis = $0.04$), suggesting a balanced distribution centered around the midpoint of the scale.

\begin{figure*}[t]
    \begin{subfigure}{0.49\textwidth}
        \centering
        \includegraphics[width=\textwidth]{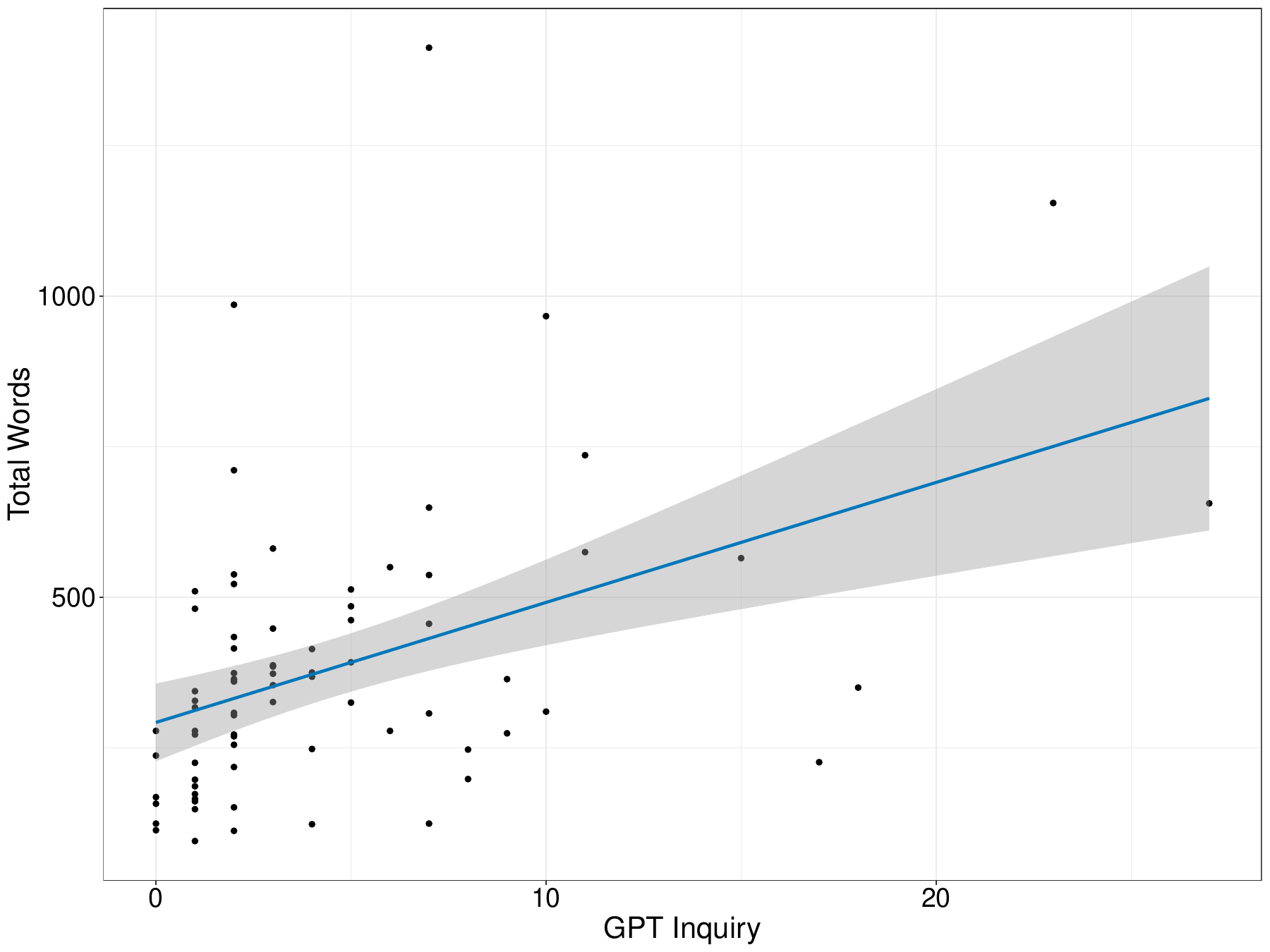}
        \caption{Word count correlation ($\rho = 0.485$, $p < 0.001$)}
        \label{fig:spear_word}
    \end{subfigure}
    \hfill
    \begin{subfigure}{0.49\textwidth}
        \centering
        \includegraphics[width=\textwidth]{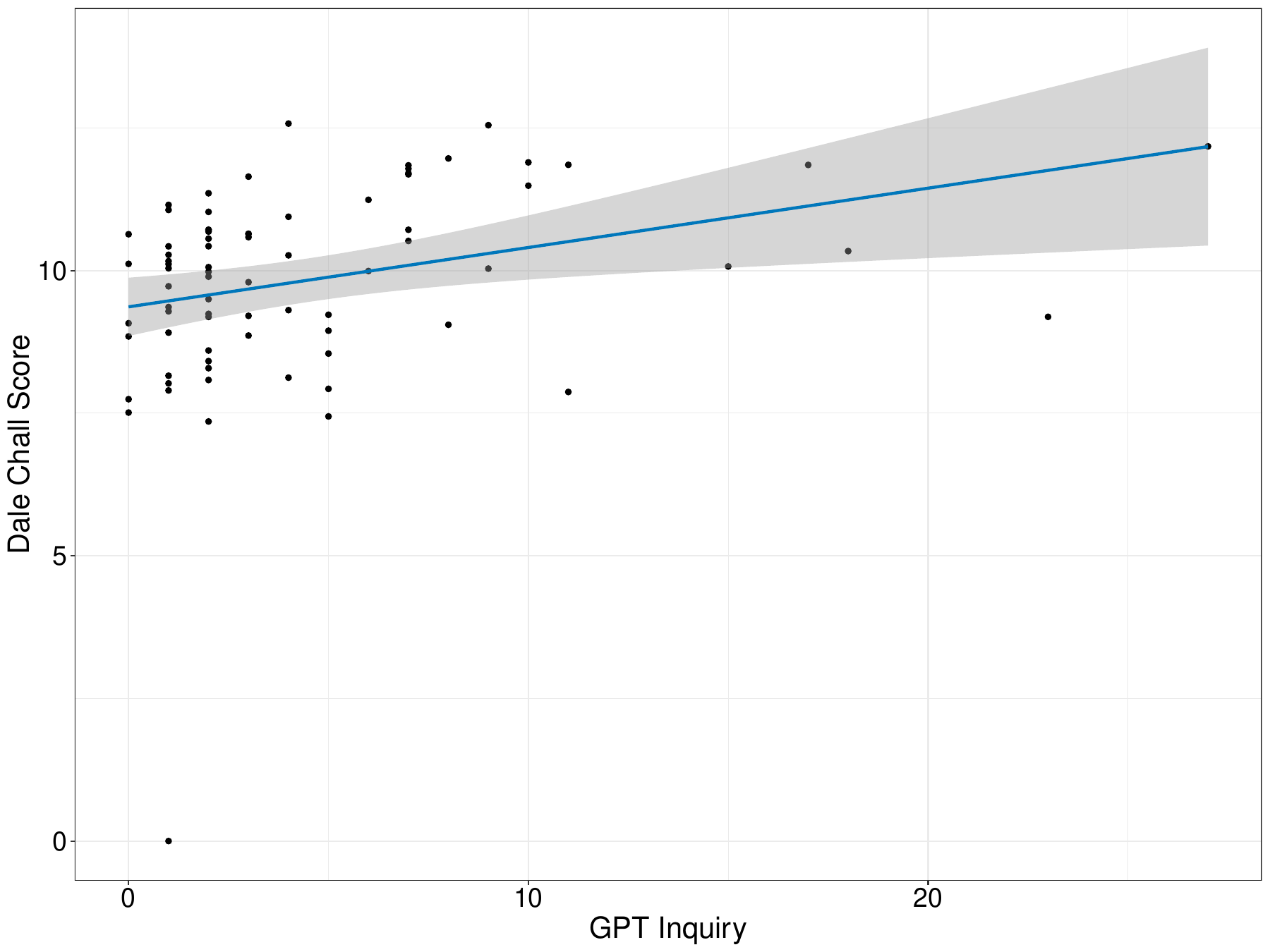}
        \caption{Dale-Chall Spearman Correlation ($\rho = 0.380$, $p < 0.001$)}
        \label{fig:spear_dc}
    \end{subfigure}
    \caption{Spearman Correlation for word count and Dale-Chall Readability}
\end{figure*}

\section{Quantitative Characterization of AI Writing Patterns} 

To demonstrate the dataset's analytical potential, we present two illustrative analyses: one examining the association between query frequency and writing outcomes (\ref{sec:querycount}, and another identifying writer profiles from patterns of AI contribution and human editing (\ref{sec:HCR-HER-definition}-\ref{sec:clustering}).

\subsection{Query Count and Writing Experience}
\label{sec:querycount}
We examined whether AI querying frequency—that is, the number of questions participants asked ChatGPT—was associated with the writing process/outcome. Given that query count is a skewed count variable, we conducted Spearman rank correlations between query count and various metrics: word count, time taken, readability, perceived ownership (PO), and the Creativity Support Index (CSI). The number of queries submitted to ChatGPT was significantly correlated with word count, time taken, and readability. Specifically, query count was positively associated with word count ($\rho = 0.485$, $p < 0.001$) and time taken ($\rho = 0.493$, $p < 0.001$), indicating that participants who asked more queries tended to produce longer essays and spend more time on the task (see \autoref{fig:spear_word}). A similar pattern was observed for readability as measured by the Dale–Chall score ($\rho = 0.380$, $p < 0.001$), with higher query counts associated with essays that are more difficult to read (see \autoref{fig:spear_dc}). No significant correlations were found between query count and the post-task survey measures (PO and CSI).

\begin{figure*}[ht!]
    \begin{subfigure}{0.49\textwidth}
        \centering
        \includegraphics[width=\textwidth]{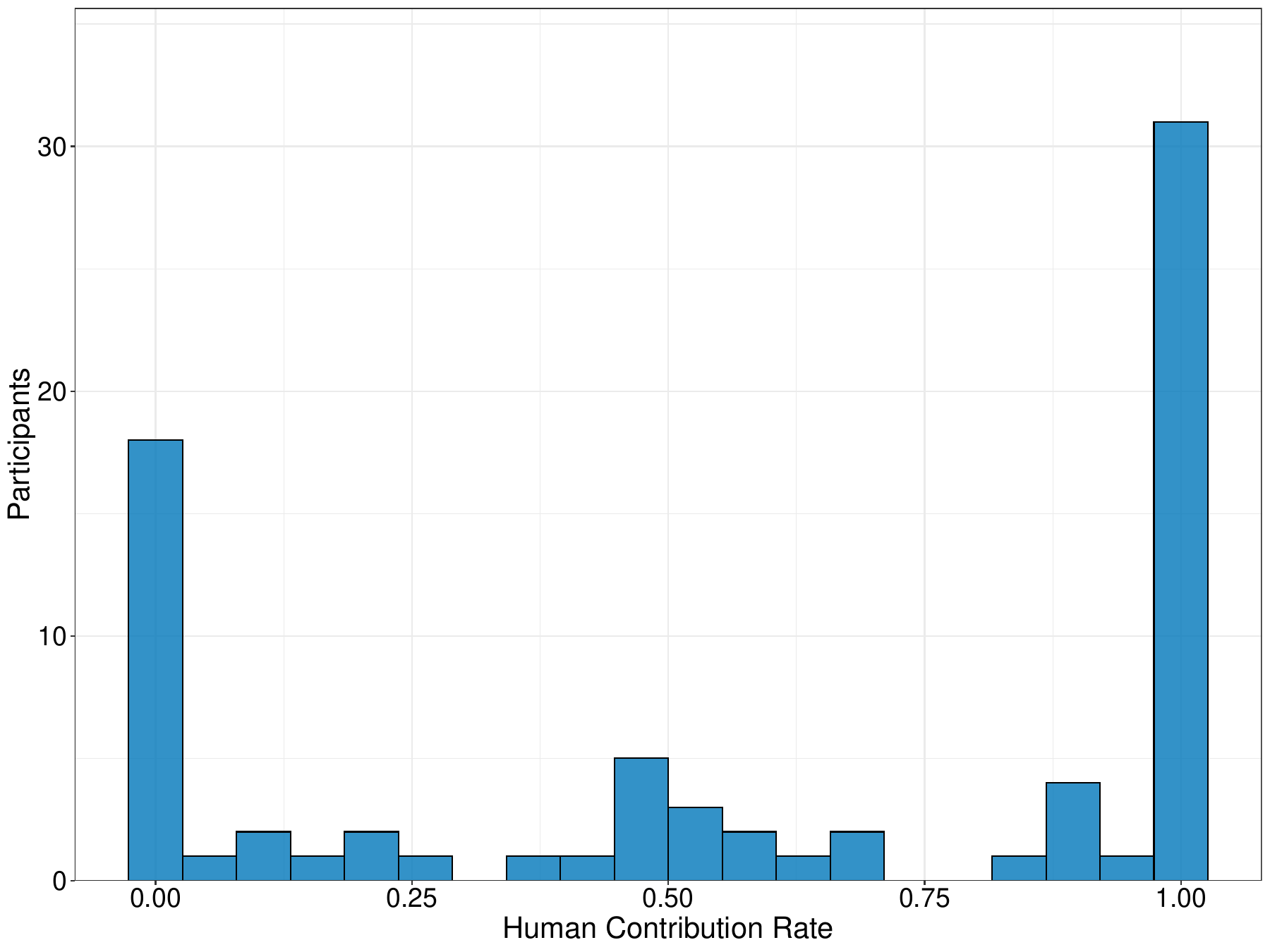}
        \caption{Distribution of the Human Contribution Ratio}
        \label{fig:HCR}
    \end{subfigure}
    \hfill
    \begin{subfigure}{0.49\textwidth}
        \centering
        \includegraphics[width=\textwidth]{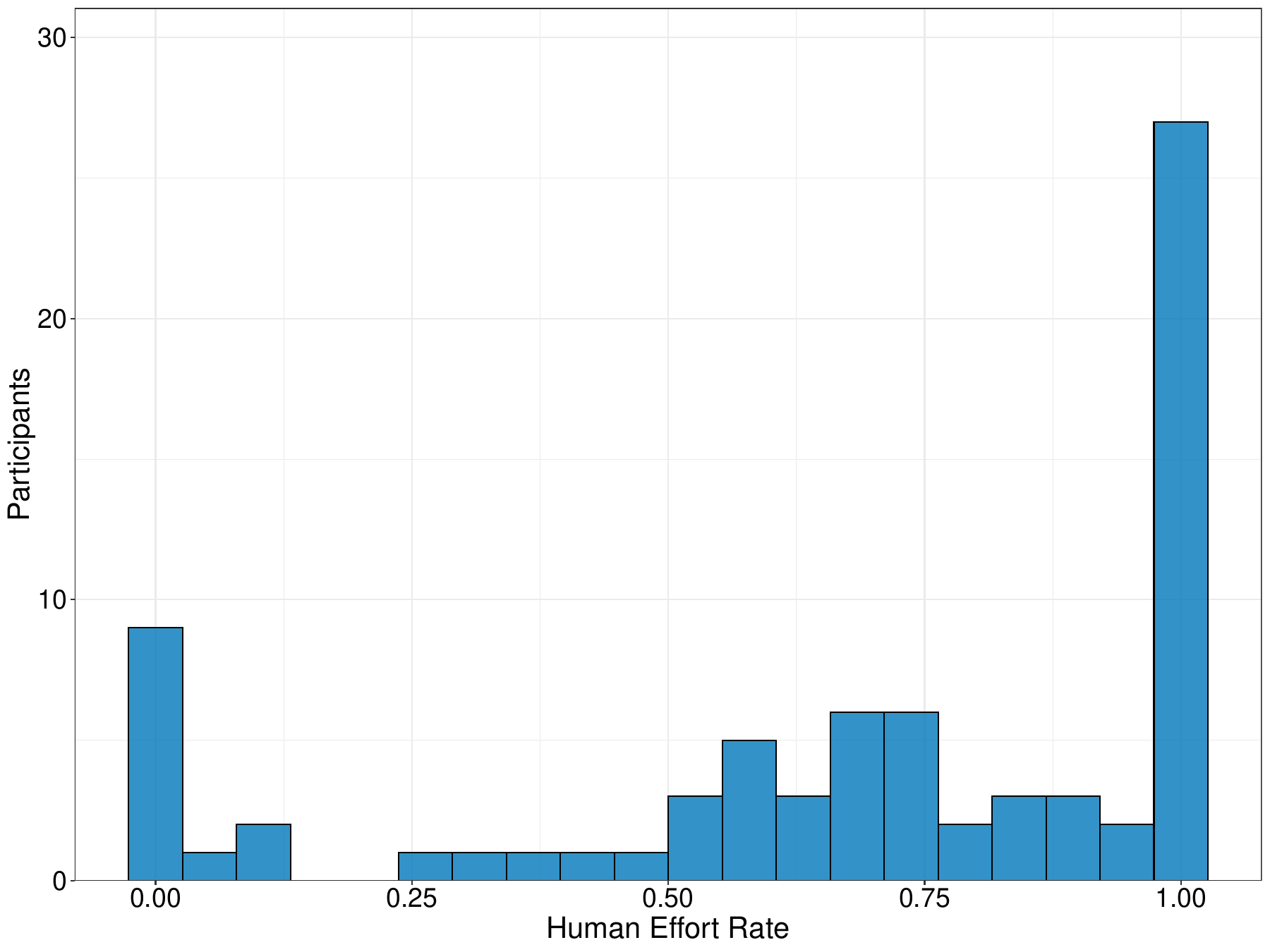}
        \caption{Distribution of the Human Edit Ratio}
        \label{fig:HER}
    \end{subfigure}
    \caption{Distributions of HCR and HER}
    \vspace{-10pt}
\end{figure*}

\subsection{Quantifying Human Contribution in AI-Assisted Writing}
\label{sec:HCR-HER-definition}

To capture the degree in which the user actively contributed to the writing process, two metrics were calculated: the Human Contribution Ratio (HCR) and the Human Edit Ratio (HER). HCR is used to determine the proportion of user-written words in the final essay. We calculated these rates based on the four numbers that the editor tracked: 

\begin{itemize}
    \item HA = the number of words that human added to the essay
    \item HD = the number of words deleted from the essay that were originally written by the human
    \item GP = the number of words pasted from a ChatGPT response
    \item GD = the number of words that the human deleted from the essay that were originally pasted from ChatGPT
\end{itemize}

\noindent The total length of the essay will be $(\text{HA} - \text{HD}) + (\text{GP} - \text{GD})$, and HCR can be calculated as: 

\begin{equation}
    HCR = \frac{\text{HA} - \text{HD}}{(\text{HA} - \text{HD}) + (\text{GP} - \text{GD})}
\end{equation}

\noindent This metric represents the proportion of text written by the participant in the final essay, ranging from 0 to 1 (inclusive). For example, an HCR of 0 indicates that the essay consists entirely of text pasted from ChatGPT, whereas an HCR of 1 means that all words in the essay were written by the participant. Figure~\ref{fig:HCR} presents the distribution of HCR scores, showing that most users are concentrated at the extremes (0 or 1), with relatively fewer observations in the middle range. This distribution suggests polarized writing behaviors: final essays tend to consist either entirely of the writer’s own words or almost entirely of text generated by ChatGPT.

However, HCR may overlook the extent to which the essay’s structure and ideas were originally contributed by the participant, particularly when ChatGPT is used for revision. For example, a user may write the entire essay without assistance from ChatGPT but later use it to proofread or revise the draft, replacing the original text with the revised version. In this case, the HCR would be 0 because all the words were pasted from ChatGPT’s responses, even though the underlying ideas and structure originated from the participant.

To address this limitation, we a developed complementary metric, the Human Edit Ratio (HER), defined as the proportion of all additions and deletions attributable to the participant, where deleting AI-generated content is counted as human contribution. HER is calculated using the following equation:

\begin{equation}
    HER = \frac{\text{HA} + \text{HD} + \text{GD}}{\text{HA} + \text{HD} + \text{GP} + \text{GD}}
\end{equation}

\noindent In this formulation, the only term excluded from the numerator is GP, which represents the number of words pasted directly from ChatGPT. The ratio ranges from 0 to 1: a value of 0 indicates that the participant did not write any words during the process, whereas a value of 1 indicates that no ChatGPT-generated text was pasted into the editor. Deletions—regardless of whether the text was originally generated by ChatGPT or written by the user—are counted as human contributions, as they reflect deliberate curation decisions about what to retain or remove.
\autoref{fig:HER} shows a distribution in which most participants are concentrated at higher HER values (above $0.75$), with a small number near 0. This pattern suggests that many students made substantive edits to their essays, whereas a subset relied heavily on AI-generated text with minimal independent contribution.

Nevertheless, HER is not without limitations. For example, it does not fully capture the cognitive effort involved in planning, ideation, comprehending ChatGPT-generated content, or deciding whether to incorporate AI-generated text—all of which may occur without substantial word-level edits. Therefore, while HER provides a more nuanced estimate of human involvement than HCR, it should be interpreted as an approximation of observable editing effort rather than a comprehensive measure of cognitive contribution.
% The HCR ratio should fall between 0 and 1 in most cases, where 0 indicates that the final essay contains only words pasted from ChatGPT and 1 indicates that the final essay consists entirely of user-written content. However, HCR is theoretically unbounded, as negative values can occur if the user deleted more words than they added, and values above 1 can occur if ChatGPT's net contribution is negative. 
\vspace{-10px}

\subsection{Clustering Writer's Profile Based on Contribution and Edit}
\label{sec:clustering}

\begin{figure}[t]
    \centering
    \includegraphics[width=\linewidth]{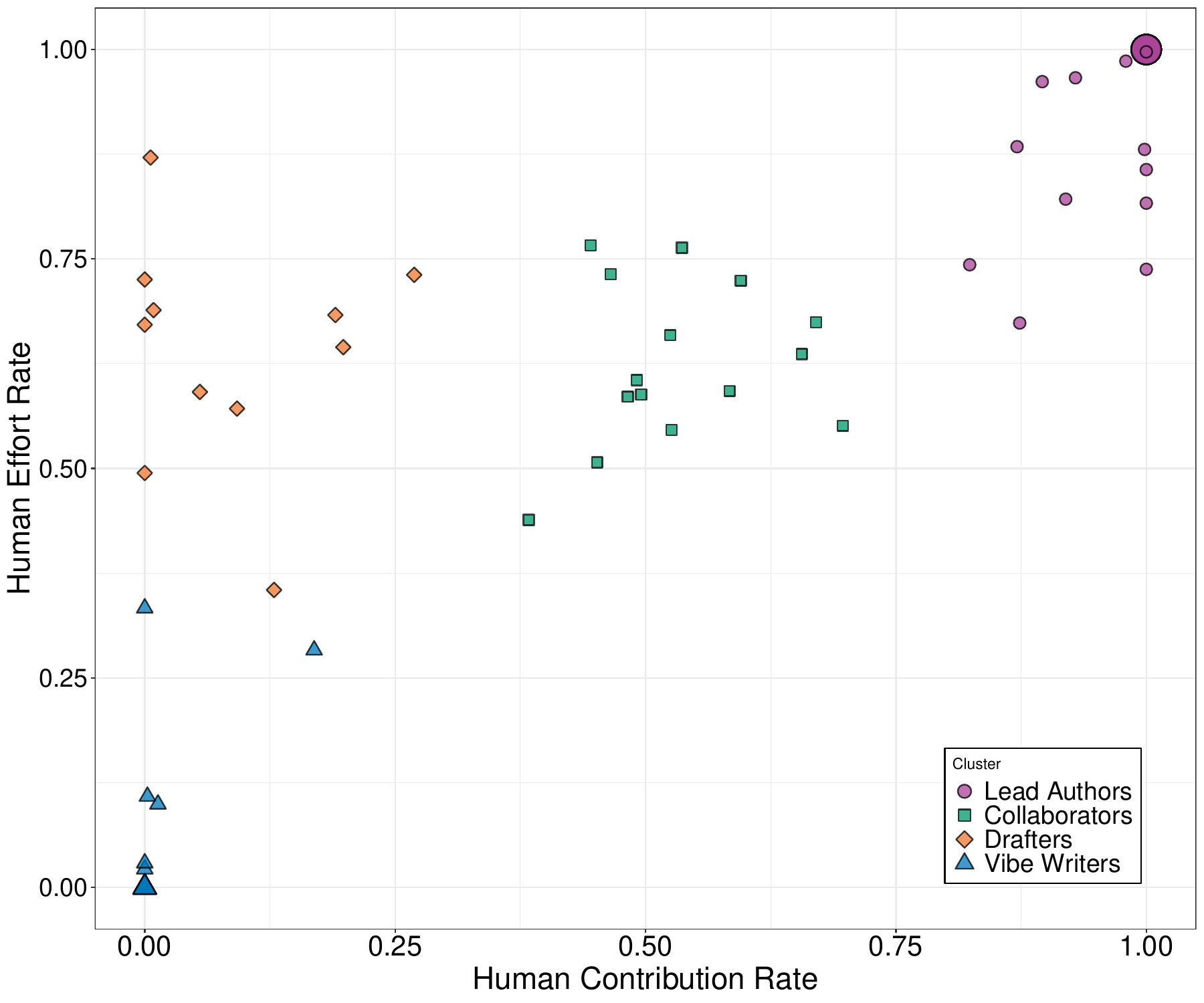}
    \vspace{-10px}
    \caption{Five Clusters of Participants based on HCR and HER. The size of the marker indicates the number of participants with the same value pairs}
    \label{fig:cluster}
    % \vspace{-10px}
\end{figure}

We used HCR and HER to identify distinct patterns of ChatGPT usage by positioning each participant within a two-dimensional space defined by the two metrics (see \autoref{fig:cluster}). We then applied K-means clustering to classify writer profiles. The optimal number of clusters was determined using silhouette analysis and the elbow method~\cite{thorndike_who_1953} based on the sum of squared errors (SSE), which indicated that $K=4$ was appropriate (see \autoref{fig:elbow} in \autoref{app:elbow}).
The resulting clusters were visualized by plotting HCR against HER and characterized according to their relative magnitudes; mean HCR and HER values are reported in \autoref{tab:clusters}. Representative essays from each cluster were manually reviewed to identify characteristic usage patterns.

The first group ($n = 37$) exhibits high HCR and high HER scores (purple circles in the top-right corner of Figure~\ref{fig:cluster}). These participants retained little to no ChatGPT-generated text in their final essays. Notably, 26 included no AI-generated words at all, suggesting they used ChatGPT primarily for idea generation, information search, or not at all. We refer to this group as \textit{Lead Authors}, as most of the writing was produced independently. In some cases, participants requested a paragraph but substantially revised it—without directly pasting the generated text—to make it their own, or used ChatGPT to explore alternative perspectives.

\begin{table}[h]
    \centering
    \caption{Average Human Contribution Ratio (HCR) and Human Effort Ratio (HER) for each cluster.(Standard Deviation).}
    \label{tab:clusters}
    \centering
    \begin{tabular}[t]{lrll>{\raggedright\arraybackslash}p{7cm}}
    \toprule
    Cluster & N & HCR & HER\\
    \midrule
    Lead Authors & 37 & 0.98 (0.04) & 0.95 (0.09)\\
    Collaborators & 15 & 0.53 (0.09) & 0.62 (0.1)\\
    Vibe Writers & 14 & 0.01 (0.05) & 0.06 (0.11)\\
    Drafters & 11 & 0.09 (0.1) & 0.64 (0.14)\\
    \bottomrule
    \end{tabular}
    \vspace{-10pt}
\end{table}

The second group ($n = 15$) is located near the center of the graph (green rectangles in the middle of Figure~\ref{fig:cluster}). These participants, whom we refer to as \textit{Collaborators}, worked alongside ChatGPT in a relatively balanced manner. Their final documents contained a fairly even mixture of ChatGPT-generated and user-written text. Typically, they generated text with ChatGPT, pasted it into the editor, and then revised it directly. In other cases, after drafting most of the document themselves, they requested a conclusion based on their existing text and incorporated it into the essay. Because these conclusions were generally summaries of content they had already written, this delegation suggests a relatively limited role for ChatGPT in shaping the overall argument.

The following group (n = 14) showed low HCR and low HER scores (blue triangles on the bottomleft cornder of Figure~\ref{fig:cluster}). These participants relied primarily on the essay generated by ChatGPT in response to the writing prompt and rarely, if ever, composed their own text in the editor. This pattern is similar to Vibe Coding~\cite{sarkar_vibe_2025}, in which a programmer uses AI to produce code and do not directly type code. Accordingly, we refer to this group as \textit{Vibe Writers}.
% ~\cite{lee_impact_2025}.

The last group, which we refer to as \textit{Drafters} (n = 11), is characterized by low HCR and high HER scores (orange diamonds on the left of the figure). This pattern suggests that although their final drafts were primarily composed of ChatGPT-generated text, they had written their own drafts at some point (as indicated by the high HER), which were later replaced with ChatGPT-generated content. In many cases, participants first wrote their own text and then asked ChatGPT to rewrite it, subsequently copying and pasting the revised version into the editor.

% \begin{figure*}[t]
%     \begin{subfigure}{0.49\textwidth}
%         \centering
%         \includegraphics[width=\textwidth]{images/cluster_inq.pdf}
%         \caption{Inquiry usage across cluster}
%         \label{fig:cluster_inq}
%     \end{subfigure}
%     \hfill
%     \begin{subfigure}{0.49\textwidth}
%         \centering
%         \includegraphics[width=\textwidth]{images/cluster_time.pdf}
%         \caption{Time taken across clusters}
%         \label{fig:cluster_time}
%     \end{subfigure}
%     \vspace{-10pt}
%     \caption{Inquiry count and time taken across clusters}
%     \vspace{-10pt}
% \end{figure*}

% \begin{figure*}[t]
%     \begin{subfigure}{0.49\textwidth}
%         \centering
%         \includegraphics[width=\textwidth]{images/cluster_dc.pdf}
%         \caption{Dale-Chall scores across cluster}
%         \label{fig:cluster_dc}
%     \end{subfigure}
%     \hfill
%     \begin{subfigure}{0.49\textwidth}
%         \centering
%         \includegraphics[width=\textwidth]{images/cluster_po.pdf}
%         \caption{Perceived Ownership across clusters}
%         \label{fig:cluster_po}
%     \end{subfigure}
%     \vspace{-10pt}
%     \caption{Readability and Perceived Ownership across clusters}
%     \vspace{-10pt}
% \end{figure*}

\subsection{Behavioral and Perceptual Differences Across Writer Clusters}

Using the previously identified clusters, we conducted a Kruskal–Wallis test to examine whether the writer groups differed significantly across measures of the writing process. Post hoc analyses were performed using Dunn’s test with Bonferroni correction to identify specific pairwise differences. Significant effects were observed for the following metrics: query count, time spent, readability, and perceived ownership.

\begin{figure}[t]
    \centering
    \includegraphics[width=0.9\linewidth, trim=10 20 10 0]{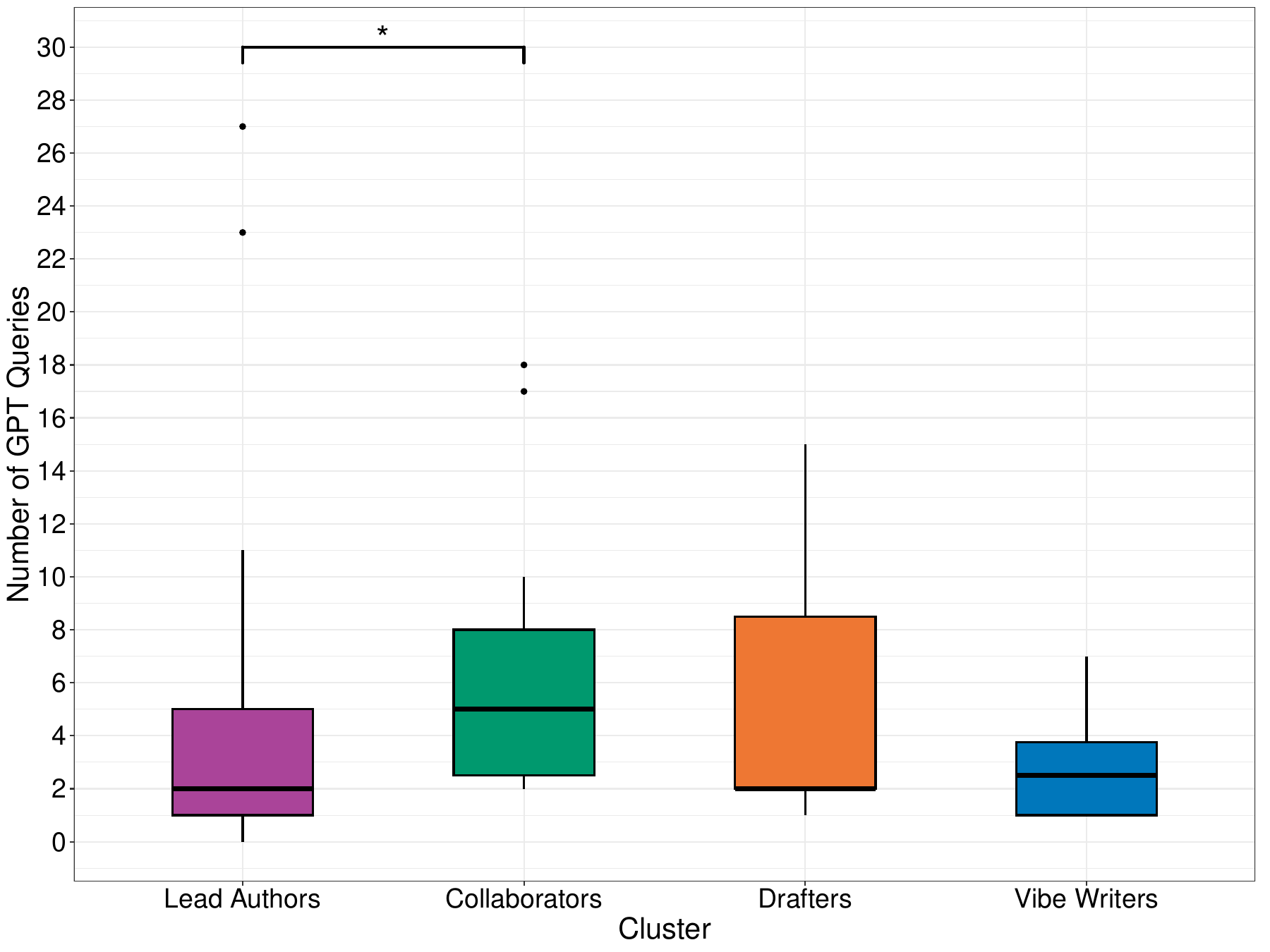}
    \caption{Inquiry usage across clusters}
    \vspace{-10pt}
    \label{fig:cluster_inq}
    % \vspace{-10pt}
\end{figure}

\begin{figure}[t]
    \centering
    \includegraphics[width=0.9\linewidth, trim=10 20 10 0]{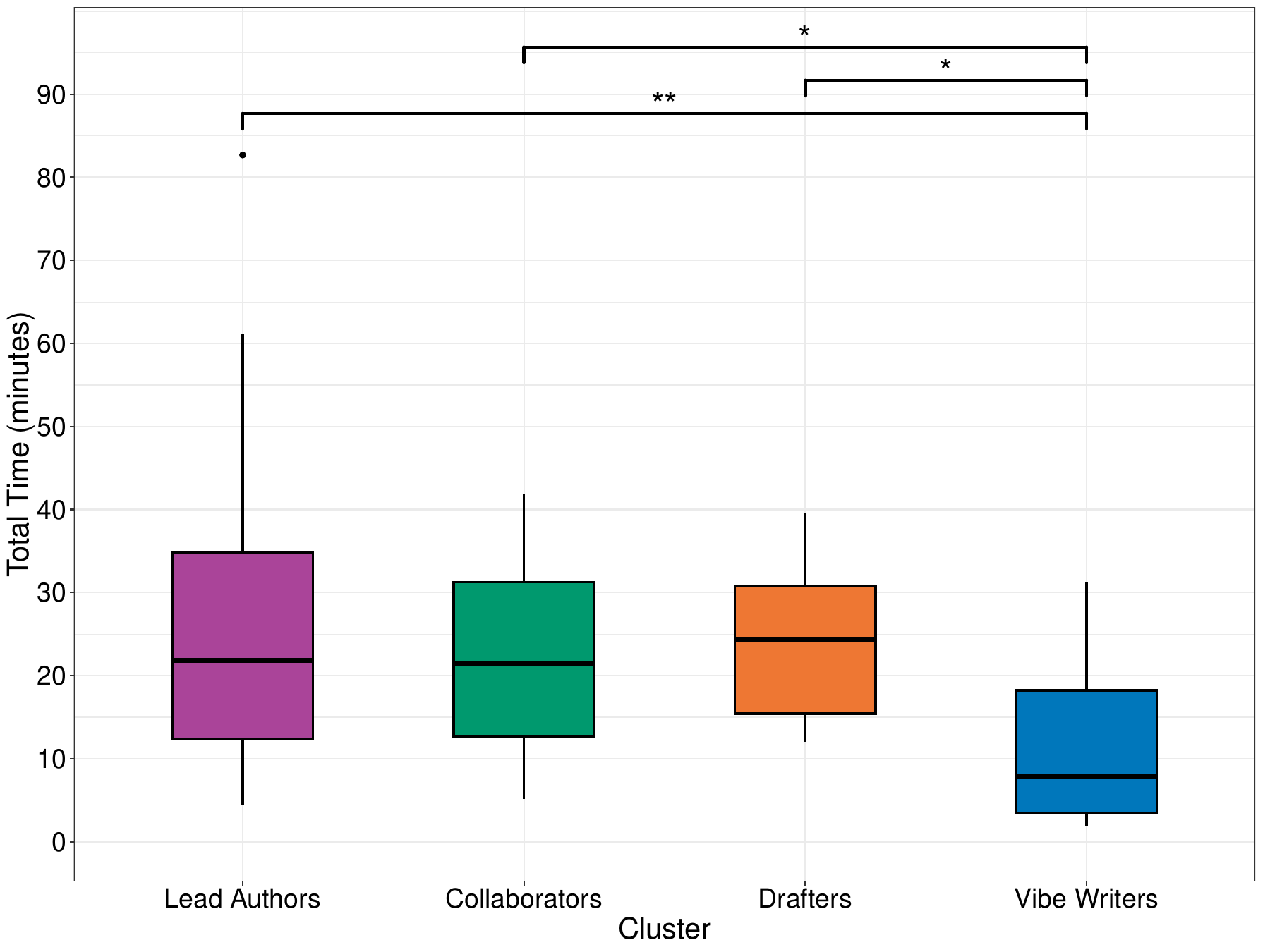}
    \caption{Time taken across clusters}
     \vspace{-10pt}
   \label{fig:cluster_time}
    \vspace{-10pt}
\end{figure}

\begin{figure}[t]
    \centering
    \includegraphics[width=0.9\linewidth, trim=10 20 10 0]{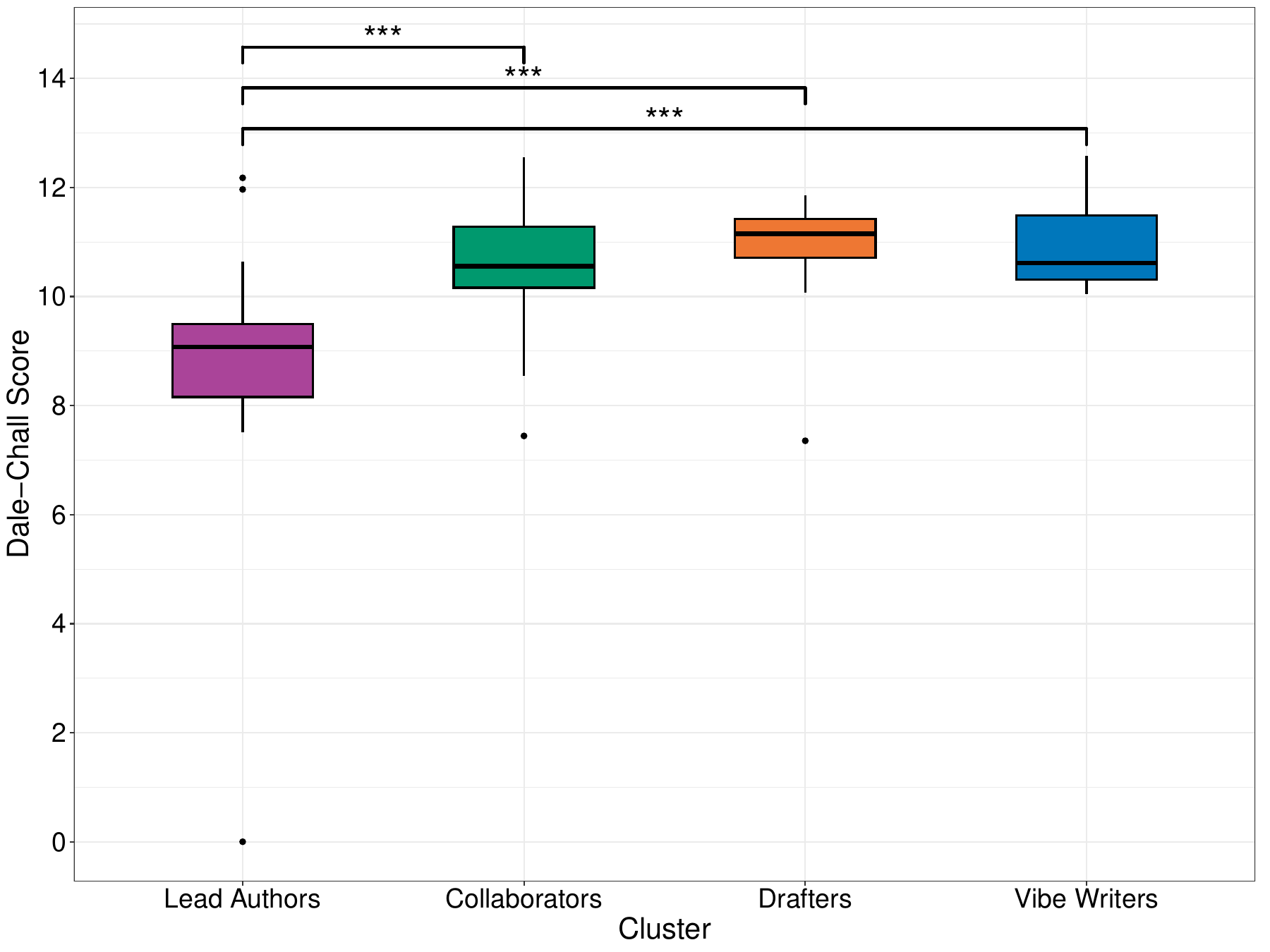}
    \caption{Dale--Chall scores across clusters}
     \vspace{-10pt}
   \label{fig:cluster_dc}
    % \vspace{-10pt}
\end{figure}

For query count, a significant difference was observed, $H(3) = 9.016, p = 0.0277$. Post hoc analysis revealed a significant difference between the \textit{Lead Authors} and \textit{Collaborators} groups ($p = 0.0141$) (see \autoref{fig:cluster_inq}), indicating that \textit{Collaborators} asked more questions than \textit{Lead Authors}.
Significant differences were also found for time spent, $H(3) = 11.69, p = 0.009$. Pairwise comparisons showed significant differences between \textit{Lead Authors} and \textit{Vibe Writers} ($p = 0.002$), \textit{Collaborators} and \textit{Vibe Writers} ($p = 0.043$), and \textit{Drafters} and \textit{Vibe Writers} ($p = 0.028$) (see \autoref{fig:cluster_time}). Vibe Writers, who generated the entire essay, completed the essay in much shorter than the other three groups. 
Significant differences were also observed in Dale–Chall scores, $H(3) = 32.26, p < 0.001$. Post hoc analysis indicated that \textit{Lead Authors} had lower Dale–Chall scores than the other three groups. These differences were statistically significant for comparisons with \textit{Collaborators} ($p < 0.001$), \textit{Vibe Writers} ($p < 0.001$), and \textit{Drafters} ($p < 0.001$) (see \autoref{fig:cluster_dc}).

\begin{figure}[t]
    \centering
    \includegraphics[width=0.9\linewidth, trim=10 20 10 0 ]{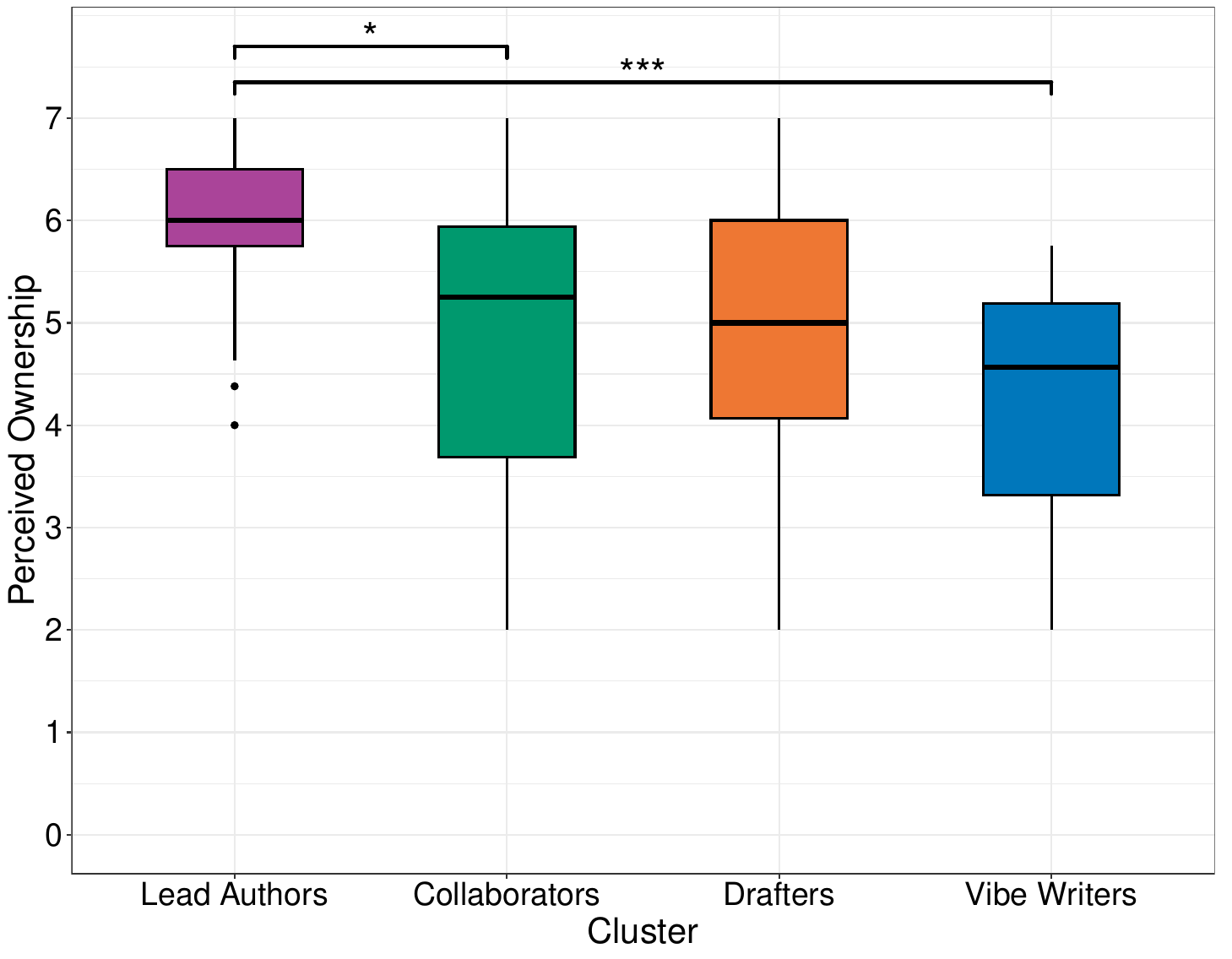}
    \caption{Perceived Ownership across clusters}
     \vspace{-10pt}
   \label{fig:cluster_po}
    % \vspace{-10pt}
\end{figure}

In the post-study survey, we measured participants’ perceptions of their writing experience using the Perceived Ownership Scale and the Creativity Support Index (CSI). A significant group difference was observed for perceived ownership, $H(3) = 23.80, p < 0.001$. Post hoc analysis indicated that \textit{Lead Authors} differed significantly from both \textit{Collaborators} and \textit{Vibe Writers} (see \autoref{fig:cluster_po}), suggesting that \textit{Lead Authors} reported stronger perceived ownership than these two groups. No significant differences were found for CSI.

\section{Replay Interface Development for In-depth Analysis}

Although this clustering provides a useful framework for characterizing usage patterns reflected in editing activity, it does not capture other important forms of interaction that do not involve direct edits. For example, P47, who belongs to the \textit{Lead Authors} group, drew ideas from ChatGPT’s response to a copy-pasted writing prompt and then composed the essay independently. Without examining the specific responses generated by ChatGPT and how those ideas were integrated into the participant’s self-written essay, it is difficult to accurately assess the extent of ChatGPT’s influence on the student’s work.
To address this limitation, we developed a replay interface that reconstructs the essay composition process, enabling more in-depth qualitative analysis. We describe the interface below and present a case study to illustrate the types of analyses this approach affords.

%In this regard, more in-depth analysis is required to understand students’ usage beyond observable editing activities; this remains an avenue for future work.

\subsection{NIRVANA: Replay System}
\label{sec:nirvana-system}

\begin{figure*}[t]
    \centering
    \includegraphics[width=0.85\linewidth, trim=10 10 10 80, clip]{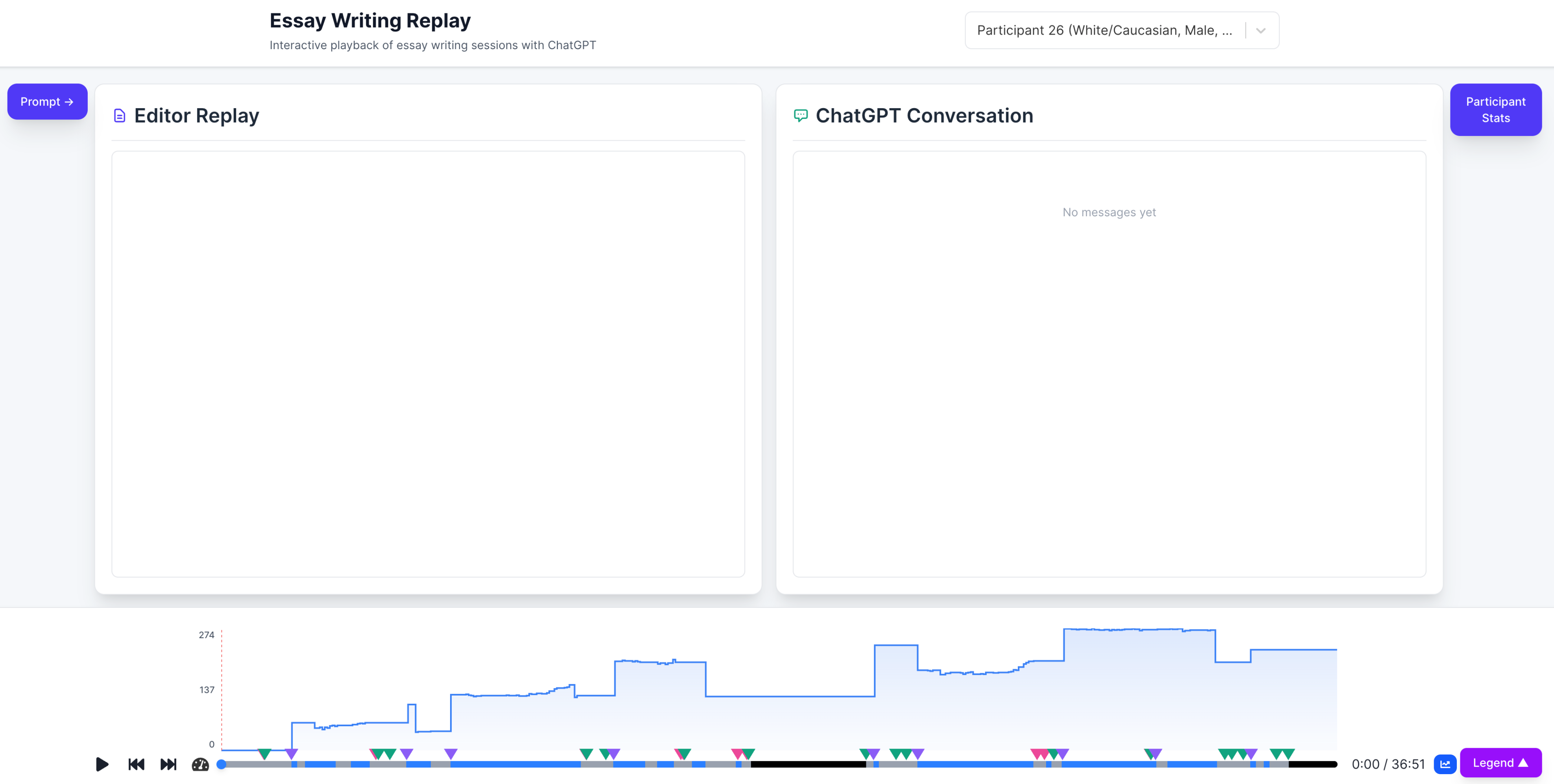}
    \caption{Writing page of our application P26's case \sang{new capture that has text. Also, give it a url.}}
    \label{fig:nirvana}
\end{figure*}

We present NIRVANA Replay System, a web-based replay system designed to make the dataset accessible and interpretable for researchers (shown in \autoref{fig:nirvana}). 

\begin{center}
    \textbf{\url{https://nirvanareplay.vercel.app/}}
\end{center}

The landing page presents all available sessions and displays summary statistics for each essay, including word count, time taken, query count, and the pre- and post-survey results, allowing researchers to identify sessions of interest before selecting an essay for review. 

Upon selecting an essay, viewers access a side-by-side replay interface showing the writing process alongside the participant’s ChatGPT queries and responses. The timeline is annotated with three event types—queries, copy events, and paste events—allowing researchers to pinpoint AI interactions and observe how generated content was incorporated. Users can pause, scrub, or step through the session to examine specific moments in detail.
A word-count-over-time graph at the bottom displays cumulative word growth, making major edits—such as pasting AI-generated text—visible as sudden increases. Together, these features support both quantitative and qualitative analyses of AI-assisted writing behavior.

% A button on the right-hand side presents the summary statistics from the landing page to reference during the writing process, and the prompt is visible through the button on the left. 

% \sang{add a graph and refer to the graph. }

\subsection{A Case Study: Instructors' Reflection}
\label{sec:case-study}

The replay interface enables trace-based (or, arguably, video-based, akin to screen recordings) qualitative analysis of writing processes, allowing researchers to examine how human–AI interactions unfold over time. Two authors of this paper—both instructors of writing-intensive courses—conducted a reflexive review of selected cases, drawing on their pedagogical experience to identify patterns they considered noteworthy or concerning with respect to students’ learning and authorship.

\noindent \textbf{P77 \href{https://nirvanareplay.vercel.app/replay?participant=77}{(Replay link)}}
P77’s case reflects explicit reliance on ChatGPT for full essay production and, based on our clustering, belongs to the \textit{Vibe Writers} group. At the outset, the student copied most of the writing prompt into ChatGPT, which produced a complete essay. The student then added a minimally constructed “personal view” in the following queries and asked ChatGPT to incorporate and contrast it with the other perspectives from the prompt, pasting those perspectives directly into the conversation. After receiving the revised output, the student pasted the generated essay into the editor without further edits.
The word-count graph, which shows a single sharp increase at the end of the session, indicates that the final essay was pasted in as a complete block of text, with only minimal priming from the student to shape the perspective. Overall, the student’s contribution was limited to framing the argument rather than engaging in the writing process itself.

\noindent \textbf{P30 \href{https://nirvanareplay.vercel.app/replay?participant=p30}{(Replay link)}}
P30’s case shows the opposite pattern: the student used ChatGPT primarily as a search engine, seeking assistance with grammar, spelling, and vocabulary; accordingly, P30 belongs to the \textit{Lead Authors} group. The only exception occurred at 25:28, when the student asked ChatGPT to make a particular sentence more concise, which may be negligible depending on the assignment context. Otherwise, most of the writing was completed independently, and the ideas were not derived from ChatGPT. The steady growth in the word-count graph over time reflects this pattern, indicating that the student took the initiative in developing the essay.

\noindent \textbf{P57: \href{https://nirvanareplay.vercel.app/replay?participant=p57}{(Replay link)}}
P57’s case is another example from the \textit{Lead Authors} group, though it reflects a different pattern of use. The participant initially asked ChatGPT to generate a complete essay based on the writing prompt. However, the prompt itself was revised, and the participant selected one of the perspectives provided in the assignment. Although ChatGPT produced a full essay, the participant used it only as a reference and did not paste the generated text into the editor. The resulting essay differed from the one generated.

\noindent \textbf{P54: \href{https://nirvanareplay.vercel.app/replay?participant=p54}{(Replay link)}}
P54 belongs to the \textit{Drafters} group. The participant initially wrote the essay independently without submitting any queries to ChatGPT. After composing a three-paragraph draft, they asked ChatGPT to elaborate further, prompting the system to propose an additional idea. The participant then requested that ChatGPT generate the proposed elaboration (e.g., “Show it to me.”), which ultimately replaced the original draft. Although the initial ideas and structure were developed by the student, ChatGPT introduced new content that P54 adopted directly into the final essay.

\noindent \textbf{P26 \href{https://nirvanareplay.vercel.app/replay?participant=p26}{(Replay link)}}
In the case of P26, the student generated the initial draft with ChatGPT and made only minimal edits to align the text with their writing style. The graphs show small, gradual revisions rather than steady linear growth (see \autoref{fig:nirvana}). Most of the text—and the underlying ideas—originated from ChatGPT. Of the three perspectives, the student adopted the one selected by ChatGPT as their own, though they may have genuinely agreed with it.
Although P26 was clustered into the \textit{Collaborators} group, qualitative analysis suggests a more concerning pattern. The student’s role was largely limited to editing, while ChatGPT acted as the primary author. This case highlights the limits of quantitative clustering and the value of qualitative analysis in distinguishing superficially similar forms of “collaboration.” It may reflect heavy reliance on ChatGPT accompanied by only surface-level revisions to present the essay as independently written.

% \sang{This is the place where we will include Dr. Maddox and Dr. Dunlap's reflection }

\section{Discussion}

In summary, our dataset provides insight into how students use ChatGPT for essay writing. As an example we conducted quantitative analysis that shows how the query count was associated with writing outcomes, including essay length, time spent, and readability. We introduced two metrics—the Human Contribution Ratio (HCR) and the Human Edit Ratio (HER)—to characterize AI integration patterns and identified four writer profiles: \textit{Lead Authors}, \textit{Collaborators}, \textit{Drafters}, and \textit{Vibe Writers}, which differed significantly in query count, time spent, readability, and perceived ownership.
To illustrate the analytical value of the NIRVANA Replay System, two authors conducted a reflexive review of selected cases, demonstrating that process-level qualitative analysis reveals differences in AI integration that quantitative clustering alone may miss. The dataset and replay system provide a foundation for human–AI interaction, learning science, and NLP research aimed at understanding how students incorporate AI into their writing. We conclude with implications and directions for future work.

% \subsection{Interpreting Query Behavior and Writing Outcomes}
\subsection{Query Behavior and Writing Outcomes}

% In our analysis, we found that query count was positively correlated with both essay length and total time taken.
% In the best case, the correlation with essay length could mean that ChatGPT acts as a generative scaffold, helping students to overcome writer's block or elaborate on ideas they may not have otherwise developed~\cite{chen_coachgpt_2025}. 
% The absence of a significant correlation between query count and CSI scores complicates this interpretation.
% Similarly, the correlation w

In our analysis, we observed a positive correlation between query count and essay length, suggesting that students who engaged more frequently with ChatGPT produced longer essays overall. An optimistic interpretation is that ChatGPT may function as a generative scaffold, helping students overcome writer’s block or elaborate on ideas they might not have otherwise developed~\cite{chen_coachgpt_2025}. Students who queried ChatGPT more often may have been more deeply engaged with the task, using the AI iteratively to develop and refine their ideas, consistent with prior research~\cite{goldi_intelligent_2024}.

At first glance, this finding may seem counterintuitive, as one might expect ChatGPT use to expedite the writing process. Indeed, the \textit{Vibe Writers} group completed the task in less time. However, this pattern may reflect limited motivation to engage deeply with the assignment rather than efficient collaboration. In this sense, intrinsic motivations (e.g., conversational engagement) and extrinsic motivations (e.g., earning a higher grade) may lead to different usage patterns. Future work should examine how incentive structures shape students’ AI use and writing behaviors.

\subsection{Understanding Writer Profiles}
% While this finding may alleviate concerns about widespread AI dependency, as the majority of students completed the work primarily on their own, it is, however, difficult to gauge the effects on learning from just these clusters alone. 

% Using K-means clustering, w
We identified four writer profiles representing distinct AI integration strategies, ranging from using ChatGPT as an ideation tool to delegating the entire writing process. Notably, the largest group, \textit{Lead Authors}, wrote their essays primarily independently despite unrestricted access to ChatGPT. This finding may alleviate concerns about widespread AI dependency, as most students submitted essays composed largely of their own text.

Behavioral differences across clusters provide further insight into these profiles. \textit{Lead Authors} produced essays with significantly lower Dale–Chall difficulty scores than the other groups, indicating more readable text aligned with their expected educational level. In contrast, essays incorporating more AI-generated content tended to exhibit greater lexical complexity and formality, characteristics often associated with large language models~\cite{klare_measurement_2000, marulli_understanding_2024}. This pattern aligns with concerns that AI-generated text may inflate surface-level complexity metrics without necessarily reflecting deeper comprehension or authentic student voice~\cite{jelson_empirical_2025}.

% The time and query patterns for the \textit{Lead Authors} group are also revealing. Despite having significantly fewer queries than \textit{Collaborators} and lower query counts on average compared to other groups, \textit{Lead Authors} spent significantly more time on the writing task overall. This suggests that their engagement with ChatGPT was selective, with more time devoted to independent writing and reflection rather than iterative querying~\cite{goldi_intelligent_2024}. In contrast, groups with higher AI incorporation completed the task more quickly, potentially indicating less time spent on independent composition and revision; rather, they delegated that task to AI.

It is important to distinguish between \textit{Drafters} and \textit{Vibe Writers}, who exhibit similarly low HCR scores but differ substantially in HER. \textit{Drafters} actively engaged in drafting before replacing their work with ChatGPT-generated content, suggesting cognitive effort that is not reflected in the final essay. This distinction raises important questions about how AI reliance should be assessed: evaluating only the final product may obscure meaningful differences in the writing process that become visible through process-level data.

\textit{Lead Authors} reported significantly higher perceived ownership than \textit{Collaborators} and \textit{Vibe Writers}, indicating that the degree of personal contribution—captured by both HER and HCR—is associated with ownership. For educators, this finding offers actionable guidance: assignments and interventions that encourage independent writing alongside selective AI use (e.g., ideation, information search, but no direct generating text) may better preserve students’ sense of authorship and agency. Rather than restricting AI access entirely, educators might focus on scaffolding how students integrate AI into their writing process.

These findings highlight the importance of moving beyond aggregate usage metrics when studying AI-assisted writing in educational contexts. The HCR and HER metrics, combined with cluster analysis, reveal a richer picture of how students position themselves relative to the AI than query count alone. Rather than asking how much students use AI, the more meaningful question for educators and researchers may be what role students assign to it in their writing process --- and whether that role supports or undermines their development as writers.

\subsection{Understanding Writing Process Through Case Analysis}

The five cases in our case study illustrate both the value and the limitations of quantitative clustering for understanding AI-assisted writing. While P77 and P30 align clearly with their respective clusters (\textit{Vibe Writers} and \textit{Lead Authors}), P26 highlights the limits of relying solely on cluster membership to characterize student engagement with AI. Although classified as a \textit{Collaborator}, P26’s process more closely resembles that of an editor refining AI-generated content than a writer actively developing ideas. This distinction is meaningful for educators concerned with learning outcomes, as editing AI-generated text engages different cognitive processes than composing independently.

The contrast between P57 and P77 is similarly instructive. Both requested full essays from ChatGPT at the outset, yet P57 used the output as reference material, whereas P77 pasted it directly without revision. This difference becomes apparent when examining how they incorporated the generated text into their writing; query-level data (e.g., questions and responses) alone would suggest a much more similar pattern of AI use.
% This behavioral difference is visible only through process-level data. This suggests that the students' intentions toward AI matter as much as the queries themselves.
% Quantitative metrics alone cannot distinguish between a student who requests an essay to learn from and one who requests it to submit.

% From an educator's perspective, the replay system provides a practical tool for understanding student behavior without requiring direct observation during writing. Patterns such as single large pastes with no subsequent editing (P77) or heavy reliance on AI-generated structure with minimal curation (P26) can serve as red flags for further review. However, this analysis also highlights the limits of automated detection. The context matters, and instructor judgment remains essential for interpreting whether a given usage pattern impacts student learning.

More broadly, these cases demonstrate that NIRVANA enables analysis that bridges quantitative and qualitative approaches. Researchers and instructors can use quantitative metrics to identify sessions of interest and then conduct qualitative reviews to understand how students engaged with AI. Critically, the process-level visibility afforded by the replay system allows educators to distinguish between surface-level similarities in final outputs and meaningful differences in learning engagement—a distinction that raw quantitative data alone may miss. For educators navigating AI in the classroom, this capability is essential: understanding how a student arrived at an essay, not just what it contains, is key to assessing learning and providing appropriate support.

% \begin{itemize}
%     \item the younger generation 
%     \item future work - qualitative coding of each query (although we did it) 
%     \item semantic similarity of GPT responses 
%     \item more sophisticated 
%     \item we should have developed the prompt replay system but we didn't.
% \end{itemize}

% \section{Previous and Future Work} % We can change this back if/when we de-anonymize our paper and cite our past work.
\section{Limitations and Future Work}

Several limitations should be considered when interpreting our findings. First, the study employed a single ACT-style argumentative prompt, which may limit generalizability to other genres such as creative writing. Second, data were collected using GPT-3.5-turbo; usage patterns and integration strategies may differ with more advanced or differently configured models (e.g., GPT-4 or domain-specialized systems). Third, the writing task was a short, approximately 30-minute session, which does not capture longitudinal writing processes or sustained AI use across a semester. Fourth, the task was not graded and carried no real academic stakes, which may have influenced students’ motivation and engagement strategies. Finally, participants were recruited through university mailing lists and Prolific, introducing potential self-selection bias and limiting the representativeness of the sample. Accordingly, our findings and the dataset should be interpreted as exploratory and correlational rather than causal, and future work is needed to examine AI-assisted writing in more ecologically and pedagogically diverse context

The NIRVANA dataset and replay system open several avenues for future research. Our analyses focused primarily on writing processes rather than the semantic content of students’ queries. Although we identified four clusters based on HCR and HER (see \ref{sec:HCR-HER-definition} and \ref{sec:clustering}), these quantitative metrics may overlook how students use ChatGPT conceptually, as illustrated in P26’s case. Existing frameworks, such as \citeauthor{flower_cognitive_1981}’s cognitive process theory of writing~\cite{flower_cognitive_1981}, could be applied to examine how ChatGPT queries align with stages of the writing process. Developing a taxonomy of writing-related queries could support automatic classification and enable systems to adapt based on instructor-defined AI policies.

Another direction is examining how students integrate ChatGPT’s responses into their essays. For example, P57 used ChatGPT output as reference material without copying it directly, yet many ideas may still have originated from the AI. Tracing such semantic provenance remains an open challenge. NLP techniques could be used to measure semantic similarity between ChatGPT responses and final essays, enabling researchers to track how AI-generated ideas are incorporated or discarded.

Essay quality also warrants further study. While we calculated Dale–Chall readability scores, we did not assess overall writing quality. Incorporating instructor-assigned scores would allow researchers to examine how AI usage patterns relate to learning outcomes and performance, distinguishing surface-level textual features from deeper indicators of writing competence.

Finally, the dataset can serve as a testbed for developing intelligent systems that characterize students’ essays and query behaviors. It may also support educators seeking to promote responsible AI use. Writing platforms that capture LLM queries (Section \ref{sec:data-collection-system}) and enforce generative-AI policies as guardrails could scaffold learning by guiding interaction rather than simply supplying answers. Such policies could be evaluated using NIRVANA, with the replay system enabling instructors to spot-check essays for potential false positives or negatives.

\begin{acks}

\end{acks}

\bibliographystyle{ACM-Reference-Format}
\bibliography{Bib}

\appendix

\section{Appendix}

\subsection{Full List of Dataset Columns}
Here, we provide a full description of all the dataset columns.
\label{app:columns}
\begin{itemize}
    \item \textbf{Participant Sheet}
    \begin{itemize}
        \item \textbf{id} - Participant id to correspond to the essay\_num in the Essay Data sheet
        \item \textbf{Gender} - Participant gender
        \item \textbf{Age} - Participant age in age bands
        \item \textbf{Race} - Participant race
        \item \textbf{Final Text} - The final essay text
        \item \textbf{Total Time} - Total time taken to complete the essay
        \item \textbf{GPT Inquiry} - Number of inquiries asked to ChatGPT
        \item \textbf{Total Words} - Total words in the Final Essay
        \item \textbf{User Add} - Number of words added by the user
        \item \textbf{User Deleted} - Number of user added words deleted
        \item \textbf{GPT Paste} - Number of words pasted from the ChatGPT response
        \item \textbf{GPT Delete} - Number of GPT Pasted words deleted
        \item \textbf{User Final \%} - Percentage of the final essay written by the user
        \item \textbf{GPT Final \%} - Percentage of the final essay pasted from ChatGPT
        \item \textbf{External Added} - Number of words added from external sources
        \item \textbf{External Delete} - Number of external words deleted
        \item \textbf{Self Efficacy Score} - Participants self efficacy score
        \item \textbf{TAM PU} - TAM subscale Perceived Usefulness
        \item \textbf{TAM PEOU} - TAM subscale Perceived Ease of Use
        \item \textbf{TAM Overall} - Technology Acceptance Model score
        \item \textbf{Perceived Ownership} - Participants feelings of Perceived Ownership for the essay
        \item \textbf{CSI Exploration} - CSI subscale Exploration
        \item \textbf{CSI Expressiveness} - CSI subscale Expressiveness
        \item \textbf{CSI Immersion} - CSI subscale Immersion
        \item \textbf{CSI Enjoyment} - CSI subscale Enjoyment
        \item \textbf{CSI Result Worth Effort} - CSI subscale Results Worth the Effort
        \item \textbf{CSI Total} - Creativity Support Index score
        \item \textbf{fk\_reading} - Flesch-Kincaid Reading Ease Score
        \item \textbf{fk\_grade} - Flesch-Kincaid Grade Level Score
        \item \textbf{dale\_chall} - Dale-Chall Readability Score
    \end{itemize}
    \item \textbf{Essay Data Sheet}
    \begin{itemize}
        \item \textbf{essay\_num} – The index of the current essay
        \item \textbf{op\_index} – The index of the operation within the current essay
        \item \textbf{time} – The timestamp of the operation
        \item \textbf{op\_loc} – The location of the operation - editor, essay\_prompt, or gpt
        \item \textbf{op\_type} – The type of the operation completed
        \item \textbf{current\_editor} – The value of the editor post operation
        \item \textbf{add} – The text added to the editor
        \item \textbf{delete} – The text removed from the editor
        \item \textbf{selected\_text} – The text highlighted in the editor or the ChatGPT inquiry \\response
        \item \textbf{cursor\_location} – The location of the cursor before the operation
        \item \textbf{recording\_obj} – The value of the codemirror record API used for playback
    \end{itemize}

\end{itemize}

\subsection{Key of Operations}
\label{app:key}
Here we provide a key for each operation.
\begin{itemize}
    \item i - insertion event
    \item d - deletion event
    \item d\_i - deletion and insertion (replace)
    \item gpt\_inquiry - inquiry made to ChatGPT
    \item gpt\_response - response made by ChatGPT
    \item l - selection of text
    \item o - cursor movement
    \item p - paste event
    \item r - drag text
    \item x - cut text
    \item y - copy event
    \item z - undo event
\end{itemize}

\subsection{Pre-Survey Questionnaire}
\label{app:pre_survey}
\textbf{What gender do you identify as?}
\begin{itemize}
    \item Male
    \item Female
    \item Non-Binary
    \item Prefer not to disclose
    \item Other: \underline{\hspace{5cm}}
\end{itemize}

\textbf{How old are you?}
\begin{itemize}
    \item 18-24
    \item 25-34
    \item 35-44
    \item 45-55
    \item 55-64
    \item 65+
\end{itemize}

\textbf{What race/ethnicity describes you?}
\begin{itemize}
    \item American Indian or Alaskan Native
    \item Asian/Pacific Islander
    \item Black or African American
    \item Hispanic
    \item White/Caucasian
    \item Other: \underline{\hspace{5cm}}
\end{itemize}

\textbf{[Questions adapted from Self Efficacy in Writing~\cite{bruning_examining_2013}]}
\textit{Please answer the following statements with a 7-point Likert scale:}
\begin{enumerate}
    \item I can think of many ideas for my writing
    \item I can transform my ideas into written text
    \item I can think of many words to describe my ideas
    \item I can come up with many new ideas
    \item I know exactly how to organize my ideas into my writing
    \item I can spell my words correctly
    \item I can write complete sentences
    \item I can punctuate correctly, i.e., put punctuation marks such as full stops and commas in my sentences
    \item I can write grammatically correct sentences
    \item I can begin my paragraphs in the right spots
    \item I can focus on my writing for at least one hour
    \item I can ignore distractions while I’m writing
    \item I can start writing assignments quickly
    \item I can control my frustration while I’m writing
    \item I can think of my writing goals before I write
    \item I can keep writing even when it’s difficult
\end{enumerate}

\textbf{How often do you use ChatGPT for writing tasks?}
\begin{itemize}
    \item Never
    \item Once in a while
    \item About half the time
    \item Most of the time
    \item Always
\end{itemize}

\textbf{[Questions adapted from TAM~\cite{davis_perceived_1989}]}
\textit{For the following, please answer based on your usage of ChatGPT (7-point Likert scale):}
\begin{enumerate}
    \item Using ChatGPT would enable me to accomplish writing tasks more quickly
    \item Using ChatGPT increases my performance in writing tasks
    \item Using ChatGPT increases my productivity in writing tasks
    \item Using ChatGPT would enhance my effectiveness in writing tasks
    \item ChatGPT makes writing tasks easier for me
    \item I have found ChatGPT useful in writing tasks
    \item Learning to use ChatGPT would be easy for me
    \item I find it easy to get ChatGPT to do what I want it to do
    \item My interactions with ChatGPT are clear and understandable
    \item I find ChatGPT flexible to interact with
    \item It would be easy for me to become skillful at using ChatGPT
    \item I find ChatGPT easy to use
\end{enumerate}

\textit{For the following, please answer based on your usage of ChatGPT (7-point Likert scale):}
\begin{enumerate}
    \item I leverage the advanced features of ChatGPT to achieve my goals more efficiently than other students
    \item I’m often interested in trying new features
    \item I maximize the capabilities of ChatGPT
\end{enumerate}

\textbf{What is the highest degree or level of school you have completed or are currently pursuing?}
\begin{itemize}
    \item No schooling completed
    \item Some high school, no diploma
    \item High school graduate, diploma or equivalent (e.g., GED)
    \item Some college credit, no degree
    \item Trade/technical/vocational training
    \item Associate degree
    \item Bachelor’s degree
    \item Master’s degree
    \item Professional degree
    \item Doctorate degree
\end{itemize}

\textbf{Have you completed the degree specified above?}
\begin{itemize}
    \item Yes
    \item No
\end{itemize}

\textbf{What is your current major?} \underline{\hspace{5cm}}

\subsection{SSE for K-Means}
\label{app:elbow}
\begin{figure*}
    \centering
    \includegraphics[width=0.49\linewidth]{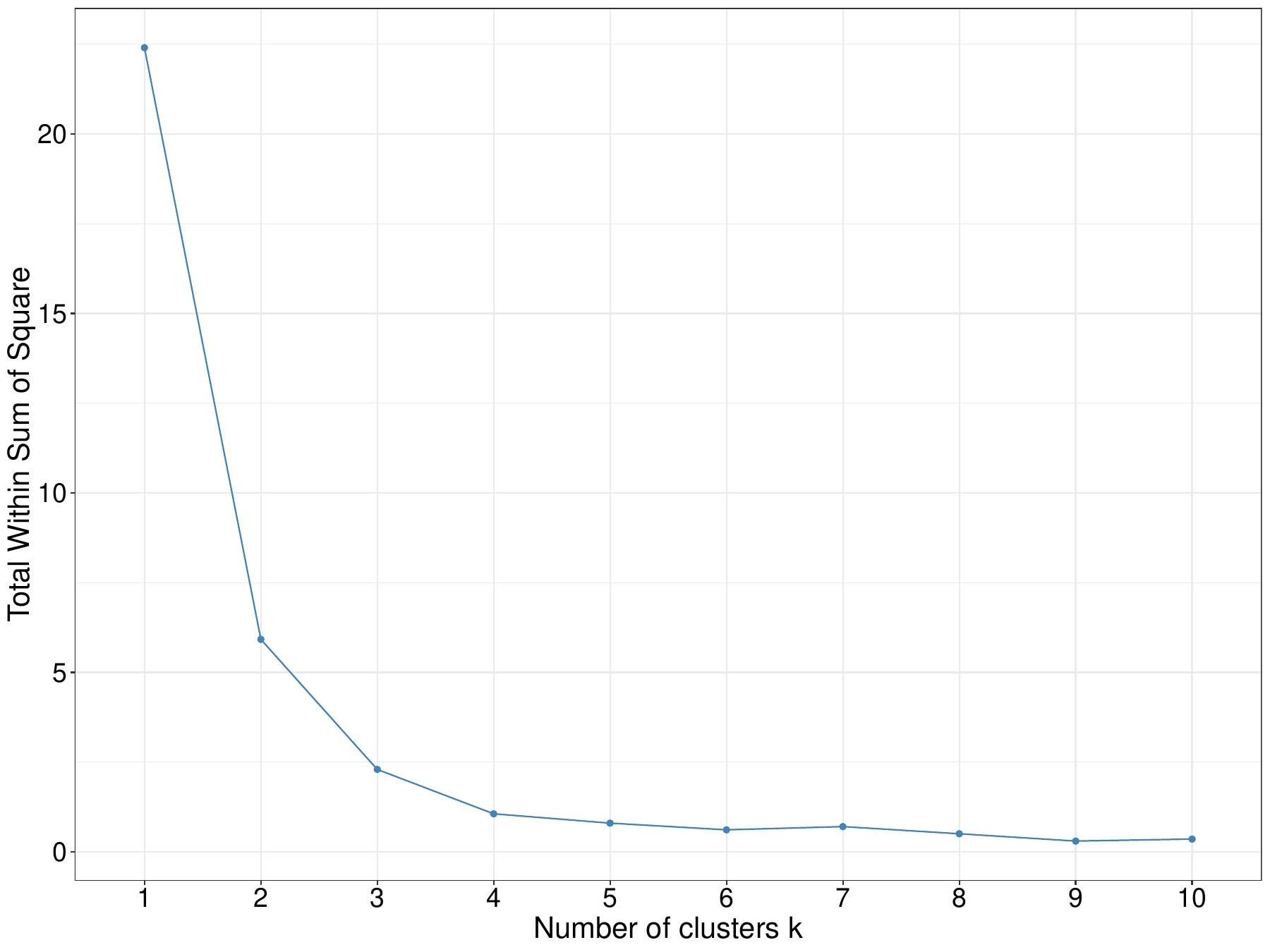}
    \caption{SSE for our clusters}
    \label{fig:elbow}
\end{figure*}

\subsection{ChatGPT page for our web interface}
\label{app:gpt}
\begin{figure*}[t]
    \centering
    \includegraphics[width=0.9\linewidth]{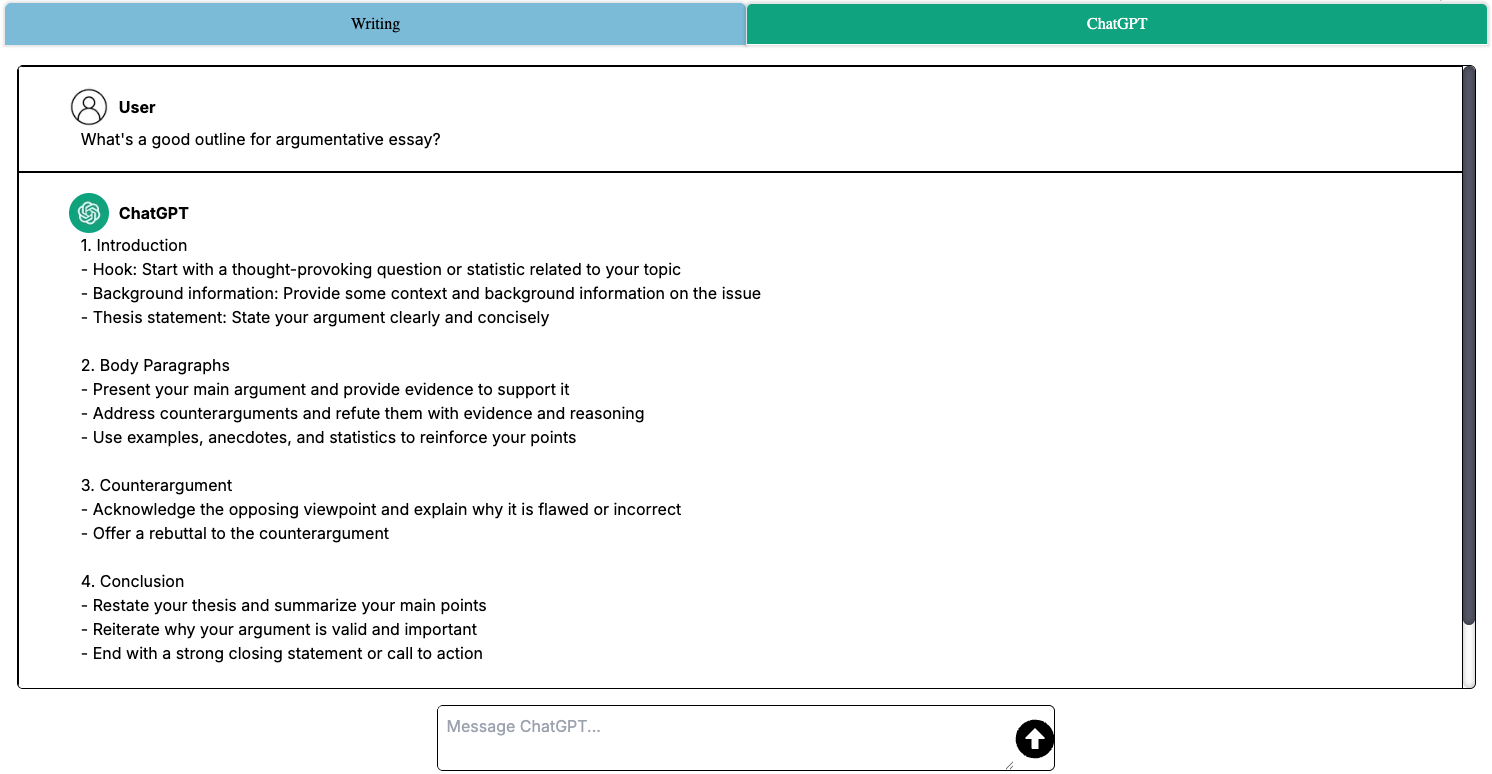}
    \caption{A screenshot of in-house ChatGPT provided to participants}
    \label{fig:gpt}
\end{figure*}

\end{document}